\documentclass{aa}

\usepackage{graphicx}
\usepackage{txfonts}
\usepackage{booktabs}
\usepackage{float}
\usepackage{comment}
\defcitealias{ThainaBatista2025}{TB25}
\def\alstar{\textsc{AlStar}}

\newcommand{\hi}{H\thinspace\textsc{i}}
\newcommand{\hii}{H\thinspace\textsc{ii}}
\newcommand{\Ha}{\ifmmode \mathrm{H}\alpha \else H$\alpha$\fi}
\newcommand{\Hb}{\ifmmode \mathrm{H}\beta \else H$\beta$\fi}
\newcommand{\nii}{\ifmmode [\mathrm{N}\,\textsc{ii}] \else [N~{\scshape ii}]\fi}
\newcommand{\oiii}{\ifmmode [\mathrm{O}\,\textsc{iii}] \else [O\,{\scshape iii}]\fi}
\newcommand{\sii}{\ifmmode [\mathrm{S}\,\textsc{ii}] \else [S~{\scshape ii}]\fi}
\newcommand{\oi}{\ifmmode [\mathrm{O}\,\textsc{i}] \else [O\,{\scshape i}]\fi}
\newcommand{\oii}{\ifmmode [\mathrm{O}\,\textsc{ii}] \else [O\,{\scshape ii}]\fi}
\newcommand{\nOne}{\ifmmode [\mathrm{N}\,\textsc{i}] \else [N~{\scshape i}]\fi}
\newcommand{\siii}{\ifmmode [\mathrm{S}\,\textsc{iii}] \else [S~{\scshape iii}]\fi}
\newcommand{\HaNii}{\ifmmode \mathrm{H}\alpha\mathrm{N}\textsc{ii} \else H$\alpha$N{\scshape ii}\fi}
\newcommand{\WHaNii}{\ifmmode \mathrm{W}_{\mathrm{H}\alpha\mathrm{N}\textsc{ii}} \else W$_{\mathrm{H}\alpha\mathrm{N}\textsc{ii}}$\fi}

\newcommand{\WHaN}{\ifmmode W_{{\mathrm H}\alpha{\mathrm N}} \else $W_{{\rm H}\alpha{\rm N}}$}
\newcommand{\WHa}{\ifmmode W_{{\rm H}\alpha} \else $W_{{\rm H}\alpha}$}

\usepackage{xcolor}
\definecolor{Jpurple}{RGB}{ 155, 0, 255 }

\definecolor{uva}{RGB}{158, 13, 83} 

\usepackage{natbib,twoopt}
\usepackage[breaklinks=true]{hyperref} 
\hypersetup{colorlinks=true,linkcolor=blue,citecolor=blue,urlcolor=blue}

\begin{document}

   \title{Analysis of spatially resolved stellar populations and emission line properties in nearby galaxies with J-PLUS}

   \subtitle{II. Results for the M51 group and first comparison with the M101 group}

   \author{J. Thainá-Batista\inst{\ref{IAA},\ref{UFSC}}, R. Cid Fernandes\inst{\ref{UFSC}}, R. M. González Delgado\inst{\ref{IAA}}, J.E. Rodríguez-Martín\inst{\ref{IAA}}, R. García-Benito\inst{\ref{IAA}}, L. A. Díaz-García\inst{\ref{IAA}}, 
   G. Martínez-Solaeche\inst{\ref{IAA}},
   D. Ruschel-Dutra\inst{\ref{UFSC}},
   V. H. Sasse\inst{\ref{UFSC}} 
\and A.~J.~Cenarro\inst{\ref{CEFCA},\ref{UA}}
\and D.~Crist\'obal-Hornillos\inst{\ref{CEFCA}}
\and C.~Hern\'andez-Monteagudo\inst{\ref{IAC},\ref{ULL}}
\and C.~L\'opez-Sanjuan\inst{\ref{CEFCA},\ref{UA}}
\and A.~Mar\'{\i}n-Franch\inst{\ref{CEFCA},\ref{UA}}
\and M.~Moles\inst{\ref{CEFCA}}
\and J.~Varela\inst{\ref{CEFCA}}
\and H.~V\'azquez Rami\'o\inst{\ref{CEFCA},\ref{UA}}
\and J.~Alcaniz\inst{\ref{ON}}
\and R.~A.~Dupke\inst{\ref{ON},\ref{MU}}
\and A.~Ederoclite\inst{\ref{CEFCA},\ref{UA}}
\and L.~Sodr\'e Jr.\inst{\ref{USP}}
\and R.~E.~Angulo\inst{\ref{DIPC},\ref{ikerbasque}}
    }
    \titlerunning{The M51 group with J-PLUS}
    \authorrunning{Thainá-Batista \& the J-PLUS collaboration}
   \institute{Instituto de Astrofísica de Andalucía (CSIC), PO Box 3004, 18080 Granada, Spain\label{IAA} \\
              \email{jullia.thainna@gmail.com}
         \and
             Departamento de Física, Universidade Federal de Santa Catarina, PO Box 476, 88040-900 Florianópolis, SC, Brazil\label{UFSC}
\and Centro de Estudios de F\'{\i}sica del Cosmos de Arag\'on (CEFCA), Plaza San Juan 1,
44001 Teruel, Spain\label{CEFCA}
\and Unidad Asociada CEFCA-IAA, CEFCA, Unidad Asociada al CSIC por el IAA y el IFCA,
Plaza San Juan 1, 44001 Teruel, Spain\label{UA}
\and Instituto de Astrof\'{\i}sica de Canarias, La Laguna, 38205, Tenerife, Spain\label{IAC}
\and Departamento de Astrof\'{\i}sica, Universidad de La Laguna, 38206, Tenerife,
Spain\label{ULL}
\and Observat\'orio Nacional - MCTI (ON), Rua Gal. Jos\'e Cristino 77, S\~ao Crist\'ov\~ao,
20921-400 Rio de Janeiro, Brazil\label{ON}
\and University of Michigan, Department of Astronomy, 1085 South University Ave., Ann
Arbor, MI 48109, USA\label{MU}
\and Instituto de Astronomia, Geof\'{\i}sica e Ci\^encias Atmosf\'ericas, Universidade de
S\~ao Paulo, 05508-090 S\~ao Paulo, Brazil\label{USP}
\and Donostia International Physics Centre (DIPC), Paseo Manuel de Lardizabal 4, 20018
Donostia-San Sebastián, Spain\label{DIPC}
\and IKERBASQUE, Basque Foundation for Science, 48013, Bilbao, Spain\label{ikerbasque}
             }

 
  \abstract
   {}
   {We aim to characterize the spatially resolved stellar population and emission-line properties of galaxies in the M51 group using the same methodology previously applied to the M101 group, ultimately seeking to understand how environmental processes shape the physical properties of galaxies by comparing results obtained for different groups.

   }
   {
   Properties are derived by applying the \textsc{AlStar} spectral fitting code to multi-band datacubes from  
   the Javalambre Photometric Local Universe Survey (J-PLUS) data.
   }
   {We present spatially resolved maps of the main stellar population and emission-line properties for the M51 group galaxies. The interacting pair M51a/b displays clearly distinct properties: M51a features prominent star-forming spiral arms, while its companion is essentially an early-type retired galaxy.
    M63 exhibits significant asymmetries in its stellar age, dust attenuation, and \Ha\ equivalent width distributions, showing signs of outside-in quenching, likely due to a past interaction. Relations between physical properties and the stellar mass surface density ($\Sigma_\star$) were investigated. The age-$\Sigma_\star$ and nebular metalicity-$\Sigma_\star$ relations are flatter than those in the M101 group. Also, unlike in the less dynamically evolved M101 group, all galaxies align with the resolved star-forming main sequence, with the exception of M51b, which has all the traits of a retired galaxy.
    }  
    {The M51 group presents signatures of more advanced dynamical evolution than the M101 group. This is evidenced by flattened age and nebular metallicity gradients, enhanced dust content, and signs of environmental quenching in some members. 
    In contrast, the less dynamically evolved M101 group largely preserves its primordial inside-out formation signatures. 
    While these findings suggest that group mass and interactions play a role in shaping galaxy evolution even in low-mass environments, the comparison of two systems is evidently limited by small-number statistics.
    This comparative study underscores the power of J-PLUS data for conducting detailed IFS-like studies of galaxies in the nearby Universe.  
    }

   \maketitle
\nolinenumbers

\section{Introduction}

Galaxy groups are key environments for studying galaxy evolution, representing an intermediate stage between isolated field galaxies and dense clusters. They are among the most common structures in the local Universe, hosting approximately half of all galaxies \citep{Eke2004}.

Groups exhibit a wide diversity of dynamical states, ranging from loosely bound associations with ongoing accretion to compact, virialized systems where interactions have significantly altered galaxy properties. These differences influence the efficiency of physical mechanisms such as mergers, tidal interactions, ram-pressure stripping, and gas inflows, which regulate star formation, chemical enrichment, and morphological transformations \citep{ZabludoffMulchaey1998, Kauffmann2004}. 
Although most studies on environmental effects in galaxy evolution rely on integrated light measurements \citep{Boselli2006}, spatially resolved analyses are essential to capture the diverse processes at play.

Star formation can be triggered or suppressed by different mechanisms, such as mergers, interactions, and ram-pressure compression, which can induce central starbursts followed by post-starburst phases \citep{Poggianti2017, Gullieuszik2020}. Conversely, ram-pressure stripping removes gas from galaxy outskirts, leading to outside-in quenching \citep{Gunn1972, Jaffe2018}. These spatially dependent effects underscore the importance of 2D mapping of galaxy properties within their environmental context.

{ 
Galaxy interactions and mergers play a central role in shaping the structural and star-formation properties of galaxies, and extensive observational and theoretical work has explored their impact across cosmic time \citep[e.g.][]{Lotz2010, Moreno2021}. However, even in the absence of strong environmental effects, central galaxies are expected to follow characteristic evolutionary pathways. In particular, both simulations and observations show that massive galaxies generally assemble their stellar mass in an inside-out way, with early, compact star formation giving way to the gradual build-up of extended stellar disks at later epochs \citep[e.g.][]{Tacchella2015, Nelson2021}. The quenching of star formation can also proceed through different spatial modes: while centrals tend to undergo inside-out quenching driven by internal mechanisms such as bulge growth or AGN feedback, satellites more commonly experience outside-in quenching associated with environmental processes \citep{Bluck2020}. 
}

This study builds on our previous work on the M101 group presented in \cite{ThainaBatista2025} ({  hereinafter \citetalias{ThainaBatista2025}}), where we have mapped
stellar population and emission line (EL) properties using {  imaging} data from the Javalambre-Photometric Local Universe Survey (J-PLUS).
Here, we {chose to} extend our analysis to the M51 group, {  a system dynamically distinct from M101, and nearly the same distance}. M51 is one of the nearest and best-studied interacting systems,
characterized by a dominant grand-design spiral galaxy interacting with its satellite companion \citep{Dobbs2010M51simulations,Tress2020_M51}, most of the galaxies present star formation activity (Fig.\ref{fig:groupM51_original};\citet{Watkins2015_M51}), and some galaxies show clear evidence of past disturbances, such as a warped \hi\ disk and stellar tidal streams \citep{Swaters2002,Tikhonov2006,Battaglia2006M63, Staudaher2015M63}. Comparing the M101 and M51 groups allows us to investigate how different interaction histories and local environments influence galaxy properties. While M101 is an unrelaxed, loosely bound group with multiple low-mass members, M51 presents an ongoing major interaction that significantly affects its stellar populations and star formation activity.

By applying a consistent methodology to both groups, we aim to examine similarities and differences in their stellar mass distributions, star formation histories, metallicity gradients, and nebular properties. This comparative approach provides a broader perspective on environmental effects in low-mass groups and contributes to the understanding of galaxy evolution in different dynamical contexts.
Future studies will extend these comparisons to other environments, making use also of the Southern Photometric Local Universe Survey \citep[S-PLUS;][]{MendesdeOliveira2019Splus}.  

The paper is organized as follows. Section~\ref{sec:data-sample} introduces the data, galaxy sample, and details the preprocessing. Section~\ref{sec:data_analysis} describes the fitting methodology with examples.
In Section~\ref{sec:results}, we present and interpret the spatially resolved maps. Section~\ref{sec:discussion} investigates the local scaling relations and presents a comparative analysis between the M51 and M101 galaxy groups.
Lastly, Sect.~\ref{sec:conclusions} summarizes our main findings and conclusions.

\section{Data, Sample and pre-processing}\label{sec:data-sample}
This study investigates the properties of galaxies in the M51 group, which includes M51, M63, NGC 5023, NGC 5229, UGC 8320, UGC 8215, UGC 8315, and UGC 8331. Following a brief overview of the data (Sect.~\ref{sec:data}) and a description of the individual galaxies in the sample (Sect.~\ref{sec:sample}), we outline the preprocessing procedures applied to the data (Sect.~\ref{sec:PreProcessing}) prior to the analysis of stellar populations and emission-line (EL) properties.

\subsection{Data}\label{sec:data}

J-PLUS uses an 80 cm robotic telescope at the Observatorio Astrofísico de Javalambre \citep{oaj} in Teruel, Spain. The Javalambre Auxiliary Survey Telescope (JAST80; \citealt{t80cam}) is equipped with a $9216 \times 9232$ pix camera (T80Cam) with a $1.4 \times 1.4\deg^2$ field of view and a pixel scale of 0.55 arcsec pix$^{-1}$. The third data release (DR3; \citealt{LopezSanjuan2024}) is used in this study. J-PLUS is distinguished by its filter system, consisting of five broad and seven narrow bands that span the $\sim 3500$-9000 \AA\ range. The data-cube was obtained with the tool \texttt{Py2DJPAS} \citep{Rodriguez-Martin2025}.
For further details on J-PLUS, including observation methodologies and data processing, see \cite{Cenarro2019jplus}, and refer to \cite{Logrono-Garcia2019}, \cite{SanRoman2019}, \cite{Lumbreras-Calle2022}, and \cite{Rahna2025} for some of the extragalactic studies using its data.

Galaxies in the M51 group are shown in Fig.\ \ref{fig:groupM51_original}. These composite images were created using the \textit{J0660} flux for the R channel, the \textit{g}-band for the G channel, and the summed fluxes of the five bluest bands (\textit{u, J0378, J0395, J0410, J0430}) for the B channel. 
EL regions appear in red in these images, as for the redshifts of the galaxies, \Ha\ and the adjacent \nii\ lines fall within the \textit{J0660} filter (centered at 6600 \AA\ and with a width of 138 \AA). The 
group is composed of two big spirals, two edge-on galaxies, and four small irregulars, with sizes from approximately 0.5 to 9 arcmin.
All the galaxies display some level of asymmetry.

\subsection{Sample}\label{sec:sample}

Table \ref{tab:sample} lists basic information on our sample galaxies. Information regarding morphologies and redshifts was extracted from NED \citep{ned1990Helou}.
We adopt a distance of 8.59 Mpc to all galaxies in the group, derived via the Tip of the Red Giant Branch method \citep{McQuinn2016M51distance}.

\begin{table}
\centering
\caption{Galaxies in our sample.}
\begin{tabular}{lccc}
\toprule[1.5pt]
Galaxy & Morphology & $\log M_\star/M_\odot$ & $M_r$ \\
\midrule[1.5pt]
M51       & interacting pair &10.79 & -21.82 \\
M51a      & SA(s)bc pec        &10.52 & -21.48 \\
M51b      & I0 pec          &10.46 & -20.42 \\
M63      & SA(rs)bc          &10.58 & -21.18 \\
NGC 5023 & Scd? edge-on      &8.75  & -17.55 \\
NGC 5229 & SB(s)d? edge-on   &8.18  & -16.23 \\
UGC 8320 & IBm               &8.08  & -16.67 \\
UGC 8331 & IAm               &7.53  & -15.18 \\
UGC 8308 & Im                &6.94  & -14.01 \\
UGC 8215 & Im               &6.67  & -13.40 \\
\bottomrule[1.5pt]
\end{tabular}
\tablefoot{The stellar mass and absolute AB r-band magnitude are derived from all spaxels of J-PLUS datacubes.
}
\label{tab:sample}
\end{table}

\begin{figure*}[h]
    \centering
    \includegraphics[width=\textwidth]{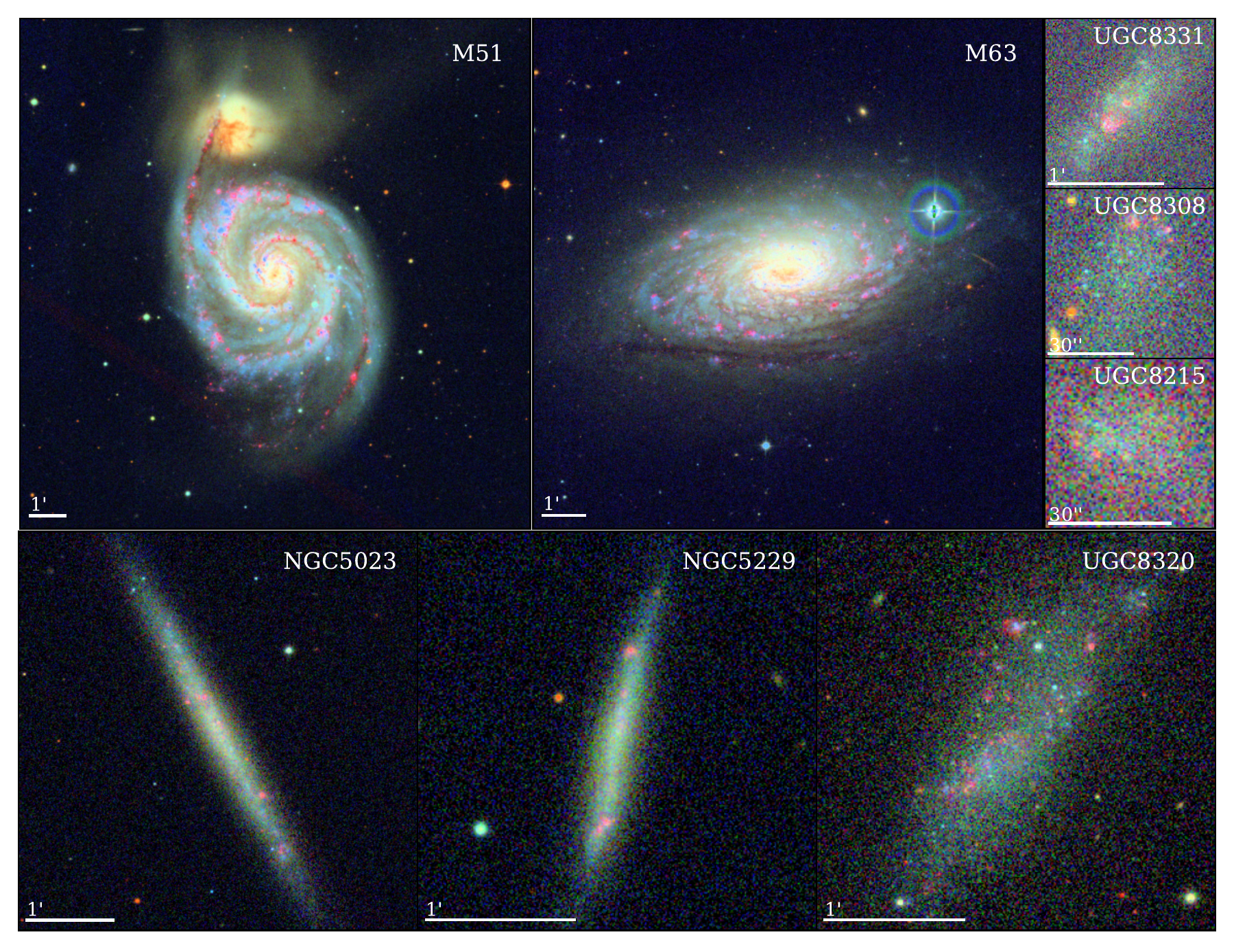}
    \caption{RGB composites using ($J0660$, $g$, sum of five bluer filters) of the original data to the galaxies of M51 group. The white line represents the distance.}
    \label{fig:groupM51_original}
\end{figure*}

{\medskip\noindent M51:}
Commonly known as the Whirlpool Galaxy, M51 comprises two interacting galaxies: The face-on, grand design spiral M51a (NGC 5194) and its early-type companion M51b (NGC 5195). With a redshift of 0.001534, M51a hosts a Seyfert 2 nucleus, plenty of star-forming (SF) regions and dust along its two arms \citep{Wei2020M51,Lee2011M51}, and a prominent tail towards M51b \citep{Watkins2015_M51}, itself a lenticular galaxy with redder colors and no recent star formation. This pair is among the most studied interacting systems, including simulations \citep{Dobbs2010M51simulations,Tress2020_M51}.

{\medskip\noindent M63:}
Also known as the Sunflower Galaxy (NGC 5055), M63 has a redshift of 0.001668, hosts a LINER nucleus, and displays flocculent spiral arms \citep{Elmegreen1987}.  
Its stellar mass exceeds that of M51a {  (see Table \ref{tab:sample})}.  
The galaxy shows prominent dust features, including nearly horizontal lanes of cold dust (visible in Fig.\ \ref{fig:groupM51_original}).  
It presents an extensive warped \hi\ disk \citep{Battaglia2006M63}.  
Observations reveal an asymmetric structure and a stellar tidal stream, likely shaped by interactions with a dwarf galaxy in the group \citep{Chonis2011M63,Staudaher2015M63}.

{\medskip\noindent NGC 5023:} 
With $z=0.001358$, this low-mass edge-on Scd galaxy exhibits an extended stellar thick disk with some \hii\ regions (Fig.\ \ref{fig:groupM51_original}), a flattened stellar halo and an \hi\ disk with some filaments at large distances \citep{Tikhonov2006}. Also known as PGC 47788 or UGC 8286.

{\medskip\noindent NGC 5229:} This edge-on galaxy with $z=0.001197$ is slightly warped. Its distorted disk and some \hii\ regions are visible in Fig.\ \ref{fig:groupM51_original}. It has a perturbed \hi\ distribution \citep{Swaters2002}. Also known as PGC 47788 or UGC 8550.

\medskip\noindent The four irregulars in the group are: UGC 8320 (also known as PGC 46039, $z=0.000640$), UGC 8331 (PGC 46127, $z=0.000867$), UGC 8308 ($z=0.000545$), and UGC 8215 ($z = 0.000726$), the smaller dwarf in the group in terms of its optical size. All they show sign of SF clumps. 


\subsection{Pre-processing}
\label{sec:PreProcessing}

\begin{figure*}
    \centering
    \includegraphics[width=\linewidth]{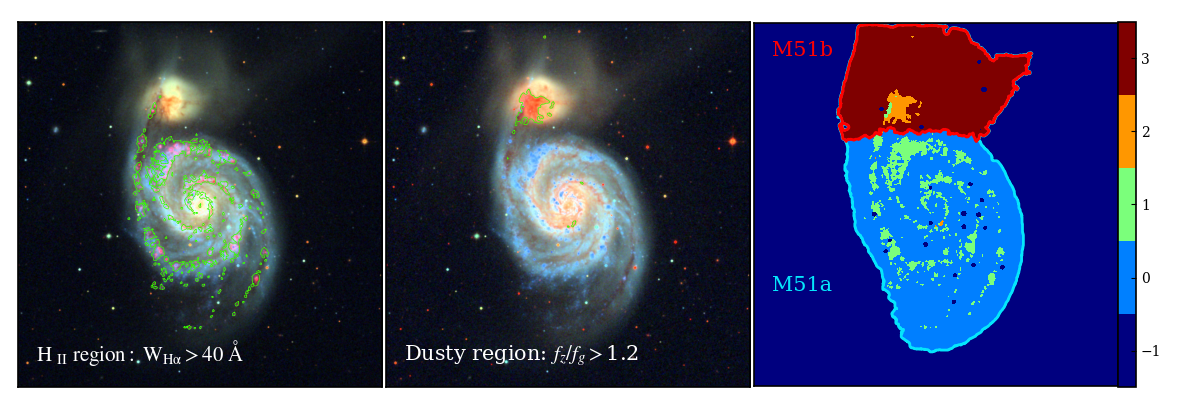}
    \caption{Illustration of the pixel tagging scheme for M51. Left panel: RGB composite 
    and green contours of the \hii\ regions detected. Middle: Same composite, but with ``dusty regions'' marked by green contours. Right: Pixel tags, where the labels $-1$, 0, 1, 2, 3 mean respectively: not galaxy data, unlabeled region, \hii\ region, dusty region, and deblended galaxy (M51b, in this case).
    }
    \label{fig:splitM51}
\end{figure*}

The data-cubes have been pre-processed to mask stars and artifacts, outline a galaxy contour, homogenize the PSF across the 12 bands, and improve its signal-to-noise by means of a combination of Butterworth filtering { \citep{Ricci2014_BW}}, $2 \times 2$ resampling, and Voronoi binning. The spatial resolution of our data after the resampling is 1.1 arcsec, corresponding to  $\sim 46$ pc. Fig.\ \ref{fig:groupM51_finalRGB} in the appendix shows how Fig.\ \ref{fig:groupM51_original} looks after these steps. These pre-processing steps were fully described in detail by \citetalias{ThainaBatista2025}, and hence will not be repeated here. We do, however, use the data on M51 to better illustrate the ``pixel tagging'' strategy devised by \citetalias{ThainaBatista2025}.

After defining a galaxy contour, masking foreground stars, applying the spatial filters, but prior to the Voronoi binning step, we tag each pixel as belonging to one of three categories: (a) \hii, (b) dusty, or (c) unlabeled. ``\hii'' pixels are identified in terms of the combined \Ha\ + \nii\ fluxes and equivalent widths estimated with the 3-Filter-Method \citep{Vilella-Rojo2015}, while ``dusty'' pixels are defined in terms of the ratio of $z$ and $g$-band fluxes, and the ``unlabeled'' tag covers all other cases; see \citetalias{ThainaBatista2025} for the exact criteria. 

The pixel tags for M51 are shown in the right panel of Fig.\ \ref{fig:splitM51}, where \hii\ (marked in the left panel), dusty (middle panel), and unlabeled tags are coded to green, orange, and blue colors, respectively. The dark red zone in this image maps an extra tag associated with the overlapping galaxy, M51b. This extra tag, not present in \citetalias{ThainaBatista2025} because no such overlapping occurs in the M101 group, was defined with the aid of the {\scshape deblending} routine in the {\scshape photutils} package \citep{LarryBradley2024_photutils}, applied to the $r$ image. 
This scheme is employed in Sect.~\ref{sec:discussion}, where we analyze M51a and M51b separately.

These tags are used in the Voronoi binning step, which is performed independently for each tag to avoid mixing physically distinct regions (approximately traced by the different tags) within the same bin. As in \citetalias{ThainaBatista2025}, the Voronoi binning is performed on the \textit{u}-band as a reference to achieve consistent signal-to-noise (S/N) across zones. The target S/N$_u$ was set to 17 for M51 and 8 for the other galaxies.

Uncertainties in the photometry ($\epsilon_\lambda$) are important for both Voronoi binning and photo-spectral fitting. Throughout large areas of our galaxies, the formal uncertainties are very small, implying, for instance, S/N over 100 in the \textit{r}-band, { while the statistical uncertainties can be <1\%, to account for systematic effects, such as flux calibration uncertainties and model limitations,that exceed these formal statistical errors, we impose an effective error floor by limiting S/N$_{r}$ to $\le 50$. So, we scaling} $\epsilon_\lambda$ in the other bands using the 
median $\epsilon_\lambda / \epsilon_r$ spectrum within the galaxy mask, hence preserving the shape of the original error spectrum. {We verified the robustness of our results by testing different S/N$_r$ ceilings (ranging from 20 to 100) and found that our main conclusions remain unchanged. The choice of $S/N_r \le 50$ was empirically selected as it yields the most physically consistent distribution of normalized residuals, $\chi_\lambda = (O_\lambda - M_\lambda) / \epsilon_\lambda$, with a dispersion closest to unity. This threshold also ensures that our Monte Carlo uncertainty propagation remains physically meaningful and is empirically supported by the resulting distribution of normalized residuals.}
For all galaxies, the errors in the seven bluest bands are $\sim 7 \times \epsilon_r$, while the four reddest bands have $\epsilon_\lambda \sim 1.4 \times \epsilon_r$.

\section{Data Analysis}
\label{sec:data_analysis}

The analysis of the J-PLUS photometry is performed with the \alstar\ code introduced by \cite{ThainaBatista2023}, with updates reported in \citetalias{ThainaBatista2025}. This section provides a summary of the code and some example fits for regions of M51 and M63.

\alstar\ decomposes photo-spectra into stellar and EL components using a non-parametric approach. The stellar base includes 80 populations, spanning 16 ages from 0 to 14 Gyr, and five metallicities in the 0.2--3.5 $Z_\odot$ range, with model spectra extracted from an updated version of \cite{Bruzual&Charlot2003} models \citep{Plat2019} for a \cite{Chabrier2003} initial mass function (IMF). The EL base contains 94 elements\footnote{The nebular base is available in \url{https://github.com/jtbatista/nebular-baseTB23}} 
with unitary \Ha\ flux and line ratios mimicking those of \hii\ regions, diffuse ionized gas (DIG), and AGN. Nebular continuum emission is accounted for \citepalias{ThainaBatista2025}.

Dust attenuation is modeled with the \cite{Calzetti2000} law, incorporating two components: dust in the diffuse interstellar medium (ISM), which affects all stellar populations and ELs, and an additional component associated with populations younger than 10 Myr to account for dust in their birth clouds (BC). ELs associated with SF regions also undergo this extra attenuation. 

The V-band optical depths associated to ISM and BC are linked by \(\tau^{\rm BC} = 1.27 \tau^{\rm ISM}\) \citep{Calzetti1994} to reduce degeneracies.
In the description of attenuation-related results, we make use of an effective optical depth, $\tilde{\tau}$, defined as:

\begin{equation}
\label{eq:tauEff}
\tilde{\tau} \equiv x_{\rm BC} (\tau^{\rm ISM} + \tau^{\rm BC}) + (1 - x_{\rm BC}) \tau^{\rm ISM}
= x_{\rm BC} \tau^{\rm BC} + \tau^{\rm ISM}
\quad,
\end{equation}

\noindent where $x_{\rm BC}$ denotes the fraction of light at 5635 \AA\ coming from populations younger than 10 Myr. This weighted average of the two components tends to $\tau^{\rm ISM}$ in the absence of young stars, and $\tilde{\tau} \to \tau^{\rm ISM} + \tau^{\rm BC}$ when $\le 10$ Myr populations dominate ($x_{\rm BC} \to 1)$.

To deal with degeneracies between \nii\ and \Ha, which are both included in the \textit{J0660} filter, an empirical prior based on the equivalent width of \nii$\lambda\lambda$6548,6584  and \Ha\ lines ($W_{\HaNii}$) is applied. Basically, the first \WHaNii\ from \alstar\ run is used to constrain the EL base elements used in a second run (see \citealt{ThainaBatista2023} for details). 

The fitting process is repeated 100 times with fluxes perturbed by Gaussian noise of amplitude $\epsilon_\lambda$. The final physical properties, such as stellar mass, age, metallicity, and EL fluxes, are derived from the averages of these Monte Carlo runs. 

\begin{figure*}[ht]
    \centering
    \includegraphics[width=0.495\textwidth]{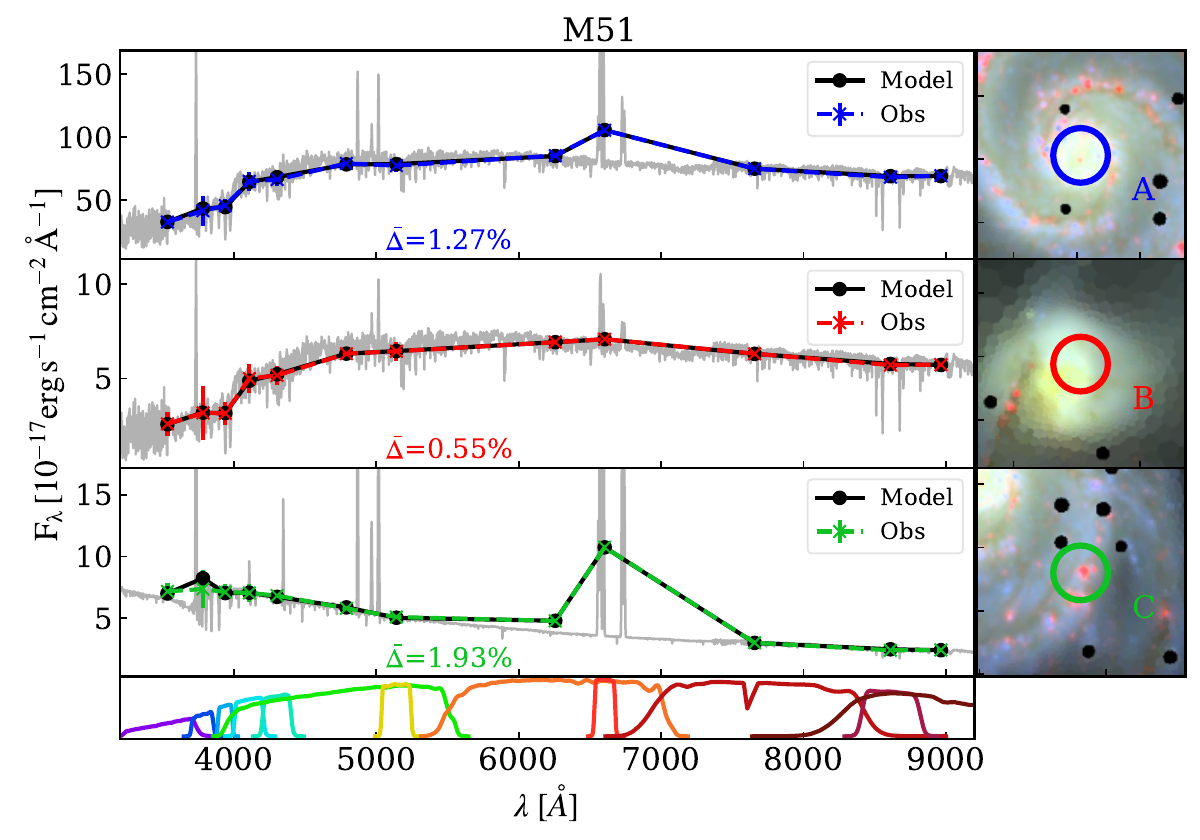}
    \includegraphics[width=0.495\textwidth]{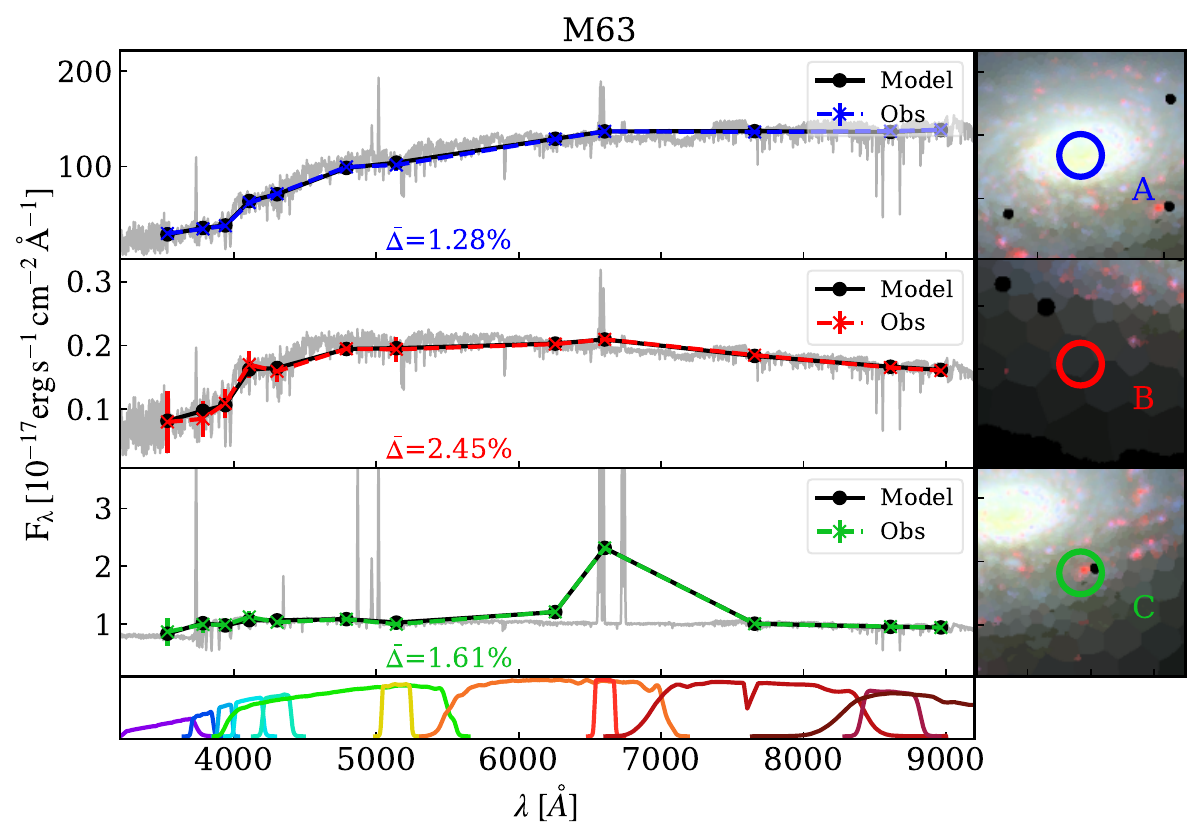}
    \caption{
    Example \alstar\ fits for individual spaxels in different regions of M51 and M63. 
    Colored lines with error bars show the data ($O_\lambda$), while black lines show the model photometric fluxes ($M_\lambda$), and the gray lines show the corresponding high-resolution model spectrum.
    Images on the right show a 1.5$^\prime\times1.5^\prime$ zoom of composites built with the \textit{J0660, r}, and \textit{g} fluxes in the R, G, and B channels, respectively. The bottom colored curves show the J-PLUS filter transmission curves.
    }
    \label{fig:fit_M51example}
\end{figure*}


Figure \ref{fig:fit_M51example} illustrates the \alstar\ fits for three regions of M51 (left) and three of M63 (right). The data are plotted with dashed colored lines and error bars, while the models are drawn in solid black. In all cases, the fluxes correspond to those inside a single $1.1^{\prime\prime} \times 1.1^{\prime\prime}$ (re-binned) spaxel, whose location is marked by a circle on the stamps on the right panels. The high-resolution spectrum associated with the \alstar\ fit is shown in gray.

The example regions in M51 are the nucleus of M51a (A), a region of M51b (B), and an \hii\ region in the disk of M51a (C). In M63, we show its nucleus (A), a faint external region (B), and an \hii\ region (C). Regions B and C in M63 are inside Voronoi zones comprising 867 and 4 pixels, respectively, while the other cases are ungrouped (re-binned) pixels. The positions of the regions are also shown in Fig. \ref{fig:positions_fitsM51eM63}.

The \alstar\ fits perform very well across all six examples, which span very different spectral shapes and brightness levels. The mean relative absolute deviation between data ($O_\lambda$) and model ($M_\lambda$) fluxes is just $\overline{\Delta} \equiv \langle |O_\lambda - M_\lambda| / M_\lambda \rangle = 1.27$, 0.55, and 1.93\% for regions A, B, and C of M51, respectively, and 1.28, 2.45, and 1.61\% for M63. Fig.\ \ref{fig:adev_violin_M51_63} in the appendix shows violin plots of the distributions of the relative residuals $(O_\lambda - M_\lambda) / M_\lambda$ for the 12 bands. In both galaxies, we see that the residuals are statistically larger in the bluer bands and smaller in the red ones, as expected from the shape of the error spectrum.

For the M51 and M63 datacubes, the median $\overline{\Delta}$ is 2.01 and 1.73\% respectively. The 16th and 84th percentiles of $\overline{\Delta}$ are (1.2, 3.2)\% for M51 and (0.9, 3.0)\% for M63. The fit quality decreases for the remaining, fainter galaxies, with median $\overline{\Delta}$ values of (in order of decreasing stellar mass) 4.25 (NGC 5023), 4.15 (NGC 5229), 3.68 (UGC 8320), 5.03 (UGC 8331), 6.67 (UGC 8308), and 5.53\% (UGC 8215). Figure \ref{fig:adev_maps_groupM51} presents the $\overline{\Delta}$ maps for the whole sample, where it is evident that the fit residuals are lower in the central, brighter regions and increase toward the fainter, outer parts of the galaxies.

\section{Results}\label{sec:results}

In this section, we present the stellar population and EL properties derived from our analysis of J-PLUS data for the galaxies in the M51 group. 
Our discussion focuses primarily on M51a/b and M63 due to their larger sizes and richer spatial structures, which allow for a detailed investigation of their maps. In contrast, the remaining group members are either significantly smaller or seen edge-on.
Although we cannot examine their internal structures in detail, we are able to characterize their global properties adequately.

\subsection{Stellar population maps}
\label{sec:StellarPropMaps_M51Group}
\begin{figure*}
    \centering
    \includegraphics[width=\linewidth]{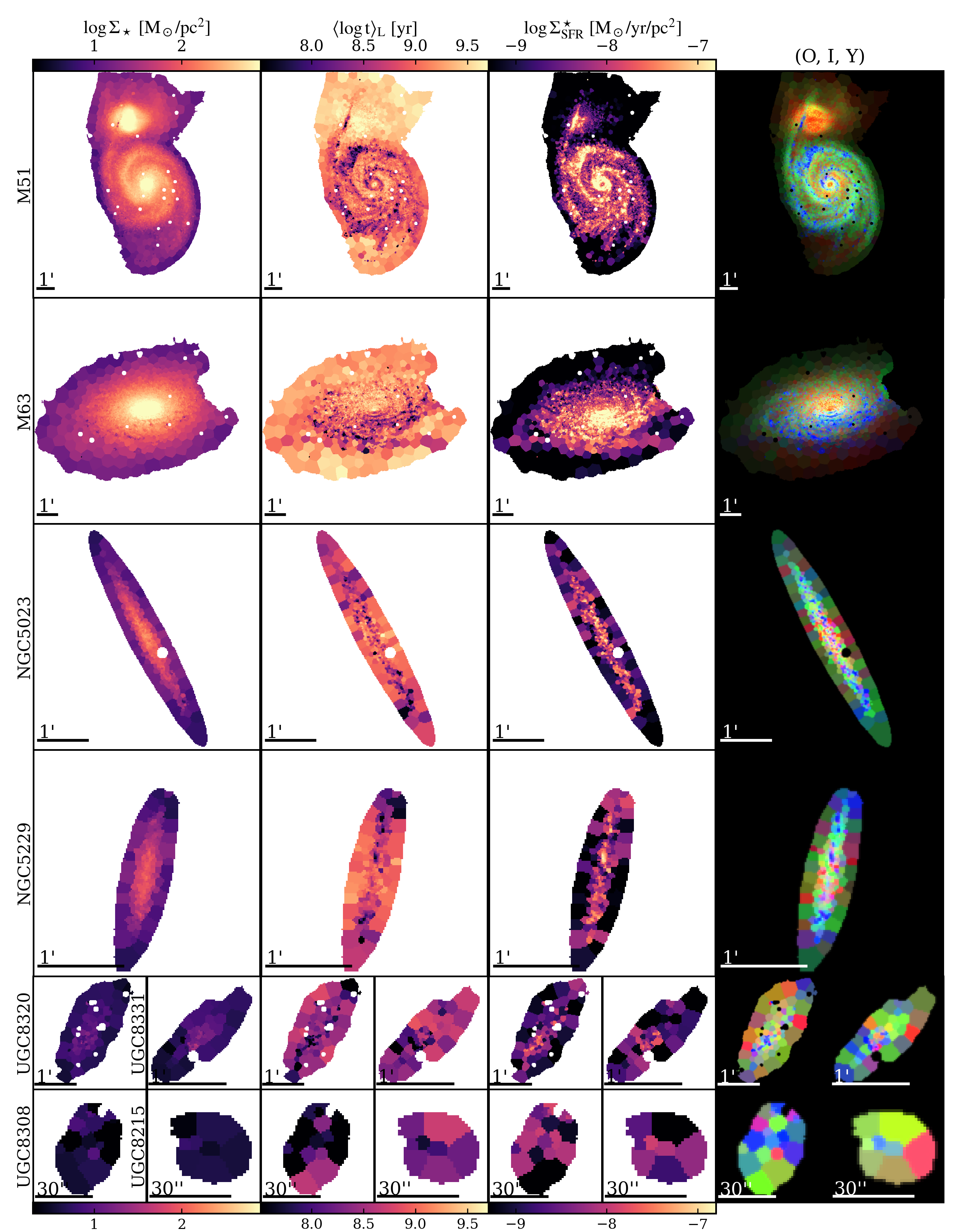}
    \caption{Maps of stellar population properties for galaxies in the M51 group. From left to right: surface density, mean age, star formation rate surface density, and an RGB with the fluxes at 5635 \AA\ of old, intermediate-age, and young populations. See text for details.}
    \label{fig:StellarPropMaps_M51Group}
\end{figure*}

\begin{figure*}
    \centering
    \includegraphics[width=\linewidth]{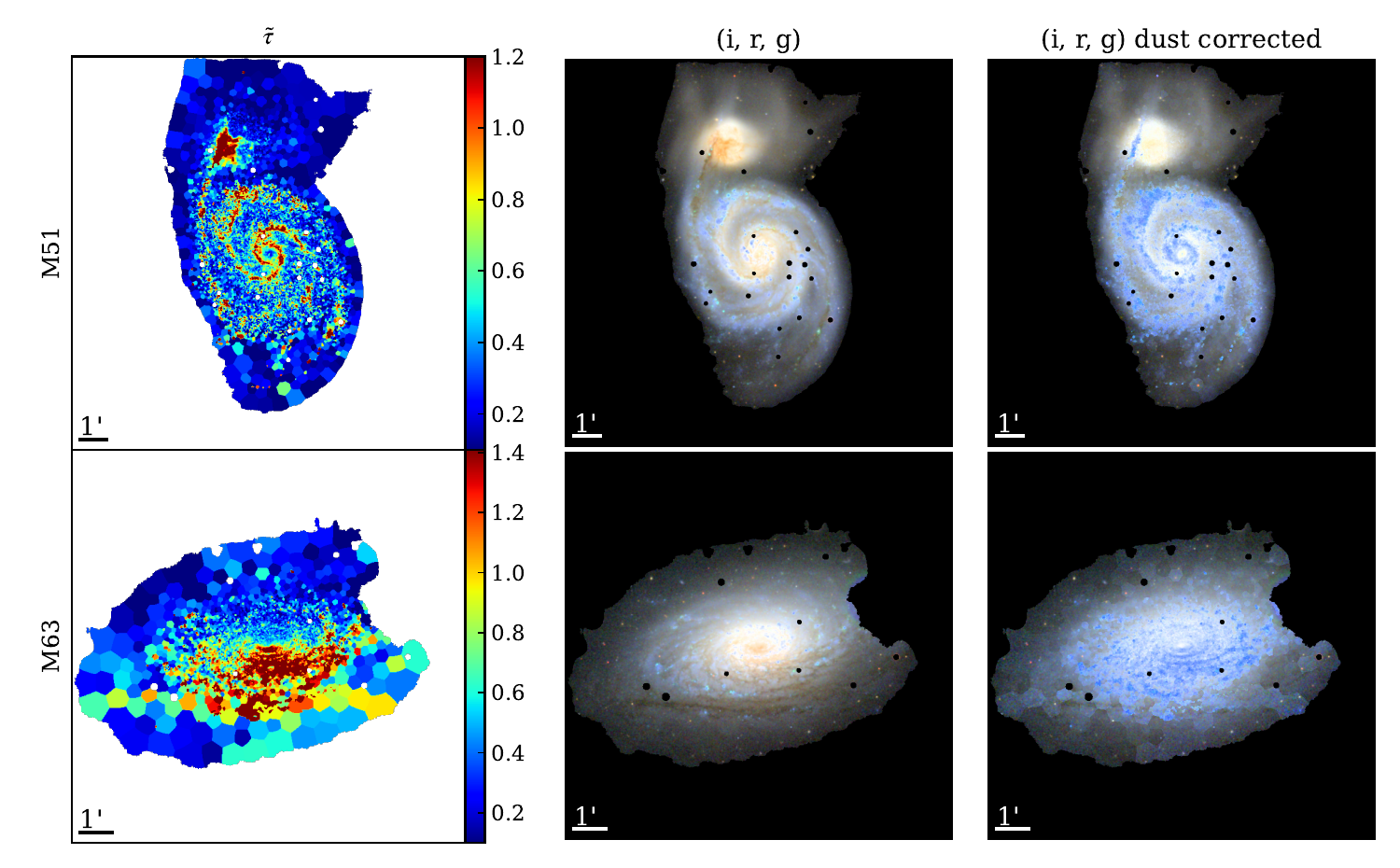}
    \caption{Left: Maps of the effective V-band dust optical depth ($\tilde{\tau}$). The central and right panels show RGB composites with the \textit{i, r}, and \textit{g} bands before and after correction for dust, respectively.}
    \label{fig:dustM51_M63}
\end{figure*}

Figure\ \ref{fig:StellarPropMaps_M51Group} presents the spatially resolved stellar population properties for the eight galaxies in the M51 group, ordered by decreasing stellar mass (top to bottom and left to right).

The stellar mass surface density $\Sigma_\star$ is shown in the first column. 
The detailed structures seen in Fig.\ \ref{fig:groupM51_original} are largely absent in the $\Sigma_\star$ images, which show smooth, centrally concentrated distributions. As is normal in spirals, young stars and dust contribute little to the mass, but have a large impact on their optical appearance.

In the case of M51a, overdensities along the spiral arms are seen in the $\Sigma_\star$ map, with arm/inter-arm ratios of the order of 2 in stellar mass density. The image also shows that M51b exhibits a more concentrated core than M51a, in agreement with the results of \cite{Wei2021M51}. The $\Sigma_\star$ peaks at approximately $13800\,M_\odot\,\mathrm{pc}^{-2}$ for M51a and $79000\,M_\odot\,\mathrm{pc}^{-2}$ for M51b, decreasing to around $6$ and $13\,M_\odot\,\mathrm{pc}^{-2}$, respectively, in the outskirts.
The total stellar masses ($\log M_\star/M_\odot$) derived for M51a and M51b are $10.52$ and $10.46$, respectively. These values are in good agreement with previous estimates, such as $10.47(10.46)$ from \cite{Wei2021M51} and $10.67(10.40)$ from \cite{MentuchCooper2012M51} for M51a(b). Given the differences in data, methodologies, and the non-trivial task of separating the two galaxies, this level of consistency is satisfactory.

M63 displays a gradient in stellar mass surface density, with central values of $\Sigma_\star \sim 3700\,M_\odot\,\mathrm{pc}^{-2}$ decreasing to $\sim 10\,M_\odot\,\mathrm{pc}^{-2}$ in the outskirts. We obtain a total stellar mass of $\log M_\star/M_\odot = 10.58$, in good agreement with previous estimates by \cite{Smith2022_M63}, who reported values of 10.54 and 10.69 using different methodologies. The lower-mass galaxies in the group show peak $\Sigma_\star$ values ranging from $210$ to $7\,M_\odot\,\mathrm{pc}^{-2}$, consistent with the expected scaling with total stellar mass \cite[figure 13 from ][]{GonzalezDelgado2014}.

The second column of Fig.\ \ref{fig:StellarPropMaps_M51Group} displays the luminosity-weighted mean log stellar age, defined as $\langle \log t \rangle_L \equiv \sum_j x_j \log t_j$,  where $x_j$ is the fraction of the flux at 5635 \AA. A first look at all galaxies reveals a  trend of decreasing mean stellar age with decreasing total galaxy mass. All galaxies display localized structures of young populations, visible as dark spots in the maps.

In M51a, prominent clumps of young stars are clearly visible along the spiral arms, with a large concentration of such regions in the side facing M51b, likely a result of enhanced star formation triggered by 
the ongoing interaction \citep{Nikola2001M51}. Notably, the northern arm extends towards and over M51b, resulting in a region where our mean ages carry contributions from stars in both galaxies.
M51b itself is dominated by old populations, as expected for an early-type galaxy, with a hint of a small inside-out drop in mean age (from $\langle \log t/{\rm yr}\rangle_L $ of about $9.76$ to $9.48$).

The age map of M63 exhibits some peculiarities. The most striking feature is the presence of internal structures of young stellar populations tracing the spiral arm pattern, while the outer regions are dominated by older populations. An intriguing result is the approximate north-south asymmetry observed in the distribution of young populations in the inner disk, with one side displaying more prominent and extended SF regions.

The low-mass galaxies in the sample consistently exhibit younger $\langle \log t \rangle_L$. The two edge-on galaxies, NGC 5023 and NGC 5229, have mean $\langle \log t/{\rm yr} \rangle_L$ over all pixels of 8.8 and 8.7. The dwarf galaxies present even younger populations, with average ages around 8.4. These lower ages are consistent with the expectations for low-mass galaxies, which tend to have more extended or delayed star formation histories compared to massive galaxies.

The third column presents maps of the recent star formation rate (SFR) surface density, $\Sigma^\star_{\rm SFR}$, derived from the \alstar\ fits. Following \cite{Asari2007_SFH}, the SFR is calculated by summing the stellar mass formed within the last $t_{\rm SF}$ years and dividing it by $t_{\rm SF}$. As in \citetalias{ThainaBatista2025}, we adopt $t_{\rm SF} = 100$ Myr.
As expected, these maps reflect features from both the $\Sigma_\star$ and $\langle \log t \rangle_L$ panels. High values of $\Sigma^\star_{\rm SFR}$ are observed not only in the central regions but also along the spiral arms.
These areas typically correspond to higher $\Sigma_\star$, consistent with the star-forming main sequence (SFMS), but are also associated with low $\langle \log t \rangle_L$ values. 
We also find that $\Sigma^\star_{\rm SFR}$ tends to decrease with decreasing total stellar mass, although this decline is less pronounced than that observed in the $\Sigma_\star$ maps.

The fourth column in Fig.\ \ref{fig:StellarPropMaps_M51Group} displays RGB composites constructed with the fluxes at 5635 \AA\ associated with young ($t \le 100$ Myr), intermediate-age ($100\,\mathrm{Myr} < t < 1\,\mathrm{Gyr}$), and old ($t \ge 1\,\mathrm{Gyr}$) stellar populations. These images can be seen as more detailed versions of the mean age maps, but with the added dimension of intensity.

In the central regions of M51a, one sees a dominant intermediate age population, surrounded by a compact bulge-like old population and then a ring of star-formation at $\sim 1$ kpc from the nucleus. From then on, blue and green regions delineate the spiral arms, the latter being visibly broader because of the longer time span in our intermediate age bin. Inter-arm regions, on the other hand, have a substantial fraction of their light attributed to old populations. The red and yellow hues in the case of M51b, indicative of its old stellar populations, are in sharp contrast with the blue-green ones of the outer part of M51a. The superposition of these two galaxies is clearly seen as the blue-green extension of M51a's arm over the reddish body of M51b.

In M63, the outer regions have practically no blue regions, suggesting a dearth of young populations in its outer disk, as previously noted in $\langle \log t \rangle_L$ map. The approximate north-south asymmetry in ages is also evident here, with blue zones being concentrated on the lower half of the image.
We have verified that in this galaxy, most of the light in our 0--100 Myr "young" bin in fact comes from $\le 10$ Myr stars. It is this very young component which drives the asymmetry in $\langle \log t \rangle_L$.

Unlike in the M101 group (\citetalias{ThainaBatista2025}), the classical inside-out color gradient is not clearly observed across the main galaxies in the M51 group. The remaining low-mass galaxies in the group do not exhibit a coherent spatial pattern in color, but generally appear dominated by younger populations.

\subsection{Dust attenuation maps}

Figure\ \ref{fig:dustM51_M63} presents the maps of the effective V-band optical depth, $\tilde{\tau}$, for M51 and M63 in the left panels, followed by RGB composites built from the $i$, $r$, and $g$ bands using both the observed (middle) and dereddened (right) fluxes. 

In M51a, high attenuation values ($\tilde{\tau} \gtrsim 1$) trace the spiral arms clearly, and in good agreement with previous studies. For example, \cite{Wei2021M51} reported $A_V$ ($1.086 \tilde{\tau}$)
values ranging from 0 to 1.2 mag using parametric SED fitting with GALEX to Spitzer data, while \cite{Wei2020M51} found a mean $A_V$ of 1.18 mag from $\Ha/\Hb$ ratios in more than 100 spectra along the arms of M51a. The highest attenuations in the M51 system are seen over M51b, where $\tilde{\tau}$ reaches values up to 3, although this dust is clearly not intrinsic to M51b itself, but from the northern arm of M51a superposed in front of it. 

For M63, we observe a similar dust distribution pattern, with enhanced attenuation ($\tilde{\tau} \geq 1.4$) tracing the spiral arms. However, a notable feature is the clear asymmetry in the dust map, which resembles the stellar age distribution seen in Fig.\ \ref{fig:StellarPropMaps_M51Group}. At first glance, this could suggest a degeneracy between age and dust (a known issue in SED fitting), but further evidence points to a more complex scenario (Sect. \ref{sec:el_WHa}).

The dereddened RGB composites for M51 and M63 appear noticeably bluer and brighter than the raw RGBs, as expected. The dust correction appears visually consistent. The obscuration of the central parts of M51b disappears nearly completely, as does the darkening along the arms of M51a.

While M51 and M63 are visibly dusty systems, the other galaxies in the group exhibit significantly lower attenuation. Across these galaxies, the average $A_V$ is approximately 0.2 mag, with mean values computed over all pixels ranging from 0.1 to 0.3 mag. This result aligns with the findings for the M101 group presented in \citetalias{ThainaBatista2025}, and is consistent with the expectation for galaxies of similar morphological types and stellar masses. 

\subsection{Emission line maps}\label{sec:ELPropMaps_M51Group}
\begin{figure*}
    \centering
    \includegraphics[width=\linewidth]{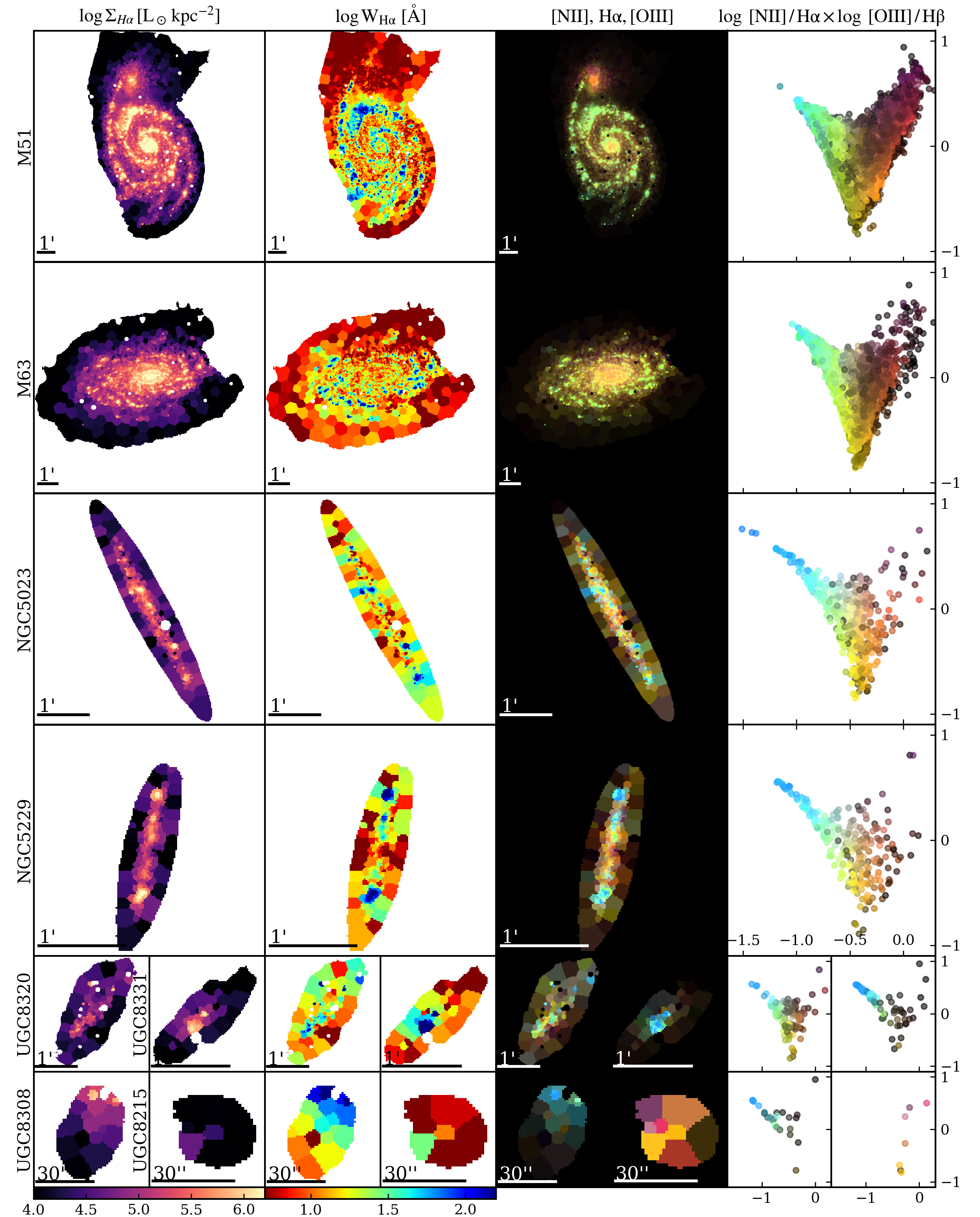}
    \caption{Maps of emission-line properties. From left to right: \Ha\ surface brightness;
    \Ha\ equivalent width;  RGB with the (\nii, \Ha, \oiii) fluxes; BPT diagram of the spaxels of each galaxy, with points color-coded as in the RGB panel. Galaxies appear in the same order as in Fig.\ \ref{fig:StellarPropMaps_M51Group}.   
    }
    \label{fig:ELPropMaps_M51Group}
\end{figure*}

We now turn to the analysis of the main EL properties derived from our fits. Fig.\ \ref{fig:ELPropMaps_M51Group} shows maps of, from left to right, the \Ha\ surface brightness ($\Sigma_{\Ha}$) and equivalent width ($W_{\Ha}$), RGB composites of the \nii, \Ha, and \oiii\ fluxes, and  BPT diagnostic diagrams ($\log \nii/\Ha$ vs.\ $\log \oiii/\Hb$; \citealt{BaldwinPTl1981}).

\subsubsection{\texorpdfstring{$\Sigma_{\Ha}$}{Sigma(Ha)} and \texorpdfstring{$W_{\Ha}$}{W(Ha)}}\label{sec:el_WHa}

The $\Sigma_{\Ha}$ maps show that, despite being an extensive property (like $\Sigma_\star$),  galaxies exhibit similar \Ha\ surface brightness levels across the group, as seen by the colorbar scale, common to all galaxies. These maps indicate that the M51 group is, in general, rich in SF regions. An exception is UGC 8215, the fainter and smaller dwarf galaxy, with a median $\log \Sigma_{\Ha}$ of $4 L_\odot/\mathrm{kpc}^{-2}$. In most galaxies, emission is concentrated in the central regions and along spiral arms, fading toward the outskirts. A similar spatial pattern is seen in $\Sigma_{\rm SFR}^\star$ (Fig.~\ref{fig:StellarPropMaps_M51Group}).
M51a is the clearest example. M51b, in contrast, shows diffuse and weak emission, similar to that seen in the inter-arm regions of M51a, with a central peak in $\Sigma_{\Ha}$ more likely due to a longer path length than to a different ionization mechanism.
In M63, the number of SF knots drops sharply in the outer part of the disk ($\gtrsim 3.7$ arcmin along the major axis), coinciding with the region devoid of young stars seen in Fig.~\ref{fig:StellarPropMaps_M51Group}. {Note that the $\Sigma_{\Ha}$ map, like all other emission-line maps presented in Fig.~\ref{fig:ELs_extrasM51group}, shows fluxes not corrected for dust attenuation.}

The $W_{\Ha}$ maps in Fig.\ \ref{fig:ELPropMaps_M51Group} are arguably more useful than $\Sigma_{\Ha}$ for diagnostic purposes, as $W_{\Ha}$ is simultaneously a proxy for the specific star formation rate (sSFR) and an indicator of the nature of the ionization source. Blue knots, corresponding to regions with $W_{\Ha}$ in excess of 100 \AA, appear in all galaxies, except in UGC 8215. 
For instance, for region C of M51a in the bottom-left panel of Fig.\ \ref{fig:fit_M51example} we obtain $W_{\Ha} = 227 \pm 18$ \AA.
These \hii\ regions match very precisely the locii of very low mean stellar age seen in  Fig.~\ref{fig:StellarPropMaps_M51Group}. This is, of course, expected, but we stress that this correspondence is not built in our fitting method, since we do not link the fit of the ELs to that of the stellar continuum.

In M51a, the spiral arms are clearly traced by the green-blue structures in the map ($W_{\Ha} \geq 25$ \AA), while in inter-arm regions $W_{\Ha} \sim 6-$19 \AA, typical of the mixed DIG as defined by \cite{Lacerda2018}. Notice also the concentration of regions of large $W_{\Ha}$ on the side of M51a facing its companion. M51b, on the other hand, has $W_{\Ha} $ around $ 5$ \AA\ throughout its body.

The $W_{\Ha}$ map of M63 shows that blue dots are confined to its inner disk. Notably, the north-south asymmetry detected in the distribution of very young populations is also present in $W_{\Ha}$, with large values found preferably in the south. 
To quantify this asymmetry we have (i) selected points with $W_{\Ha} > 30$ \AA\ and in the SF wing of the BPT diagram, (ii) split the galaxy along its major axis, and (iii) compared the mean $W_{\Ha}$ values in the two halves. 
We obtain 60 \AA\ in the northern half and 78 \AA\ in the southern one, while splitting the galaxy along the perpendicular direction yields nearly identical values (71 \AA\ in the east and  72 \AA\ in the west). 
We repeated the test using the combined $\Ha + \nii$  equivalent widths derived with the 3-filter method (which has been validated and applied in other studies, e.g. \citealt{Rahna2025} and \citealt{Lopes2025}), 
obtaining the same result. Since ELs and stellar populations are modeled independently, this confirms that the asymmetry is real and not an artifact of our fits.

The reason for this asymmetry is not evident, but, just like the profusion of SF activity in the regions of M51a closest to M51b, interactions are an evident possibility. \cite{Chonis2011M63} suggests that M63 has undergone a recent interaction with a dwarf galaxy. Additionally, \cite{Battaglia2006M63} reports a misalignment between the optical and \hi\ gas distributions. By overlaying optical and Spitzer images, we also detect signatures of cold dust in front of the galaxy, visible in the southern region of Fig.\ \ref{fig:groupM51_original}. 
{  Earlier evidence that M63 displays asymmetric age structures was reported by \cite{Sanchez-Gil2011M63M51}, however, in the east-west direction and in a different age ranges that are not directly comparable to our findings.}
Finally, we note that UGC 8308 also shows a marked asymmetry in its $\Sigma_{\Ha}$ and $W_{\Ha}$ maps, although our spatially resolved results for this dwarf galaxy are evidently limited (only 25 Voronoi zones).

\subsubsection{Emission line ratios}

Let us now examine results for other ELs, which, as demonstrated by our previous work with S-PLUS and J-PLUS (\citealt{ThainaBatista2023}, \citetalias{ThainaBatista2025}), can be adequately estimated with our methodology. The third column of Fig.\ \ref{fig:ELPropMaps_M51Group} presents composite images built from the flux maps of three key ELs: \nii\ (in the red channel), \Ha\ (green), and \oiii\ (blue). 
The fourth column shows the corresponding BPT diagram, where each spaxel is plotted with the same color it appears in the RGB composite. Together, these plots synthesize the spatial and spectral variation of ELs across the galaxies in our sample, enabling a quick visualization and assessment of the dominant ionizing mechanisms in different galactic regions.

M51a and M63 exhibit remarkably similar patterns: the spiral arms are predominantly green, indicating strong \Ha\ and comparatively weak \nii\ emission, typical of SF regions. 
In the case of region C of M51a (Fig.\ \ref{fig:fit_M51example}), for example, we find $\log \nii/\Ha = -0.62 \pm 0.15$, while for region C of M63 in the same figure $\log \nii/\Ha = -0.69\pm 0.14$. Note also that the colors of the \hii\ regions in these RGBs are similar at all radii, indicating little variation in nebular metallicity.

The EL pattern in central and inter-arm regions tends toward orange and red hues due to the increasing contribution of \nii, leading to line ratios typical of LINERs and DIG. This is even more prominent in M51b, as is evident from its red look in the RGB composites in Fig.\ \ref{fig:ELPropMaps_M51Group}.
Region B in Fig.\ \ref{fig:fit_M51example} for instance, has $\log \nii/\Ha = -0.27 \pm 0.15$ and $W_{\Ha} = 6.6 \pm 4$ \AA.

This body of evidence, combined with its weak ELs attributable to ionization by post-AGB stars \citep{Stasinska2008}, and its very old stellar populations (Fig.\ \ref{fig:StellarPropMaps_M51Group}) and early-type morphology
strongly supports its classification as a retired or quenching galaxy.
The low values of $W_{\Ha}$, with 16--84 percentiles spanning the 3--8 \AA\ range, are qualitatively consistent with this interpretation, albeit slightly above the $W_{\Ha} \leq 3$ \AA\ limit proposed by \citealt{CidFernandes2011} to identify retired systems. Taking into account that the typical (median) uncertainty in $W_{\Ha}$ for M51b is 4 \AA, as well as the evident difficulty in estimating such low equivalent widths with our data, this minor formal inconsistency should not be given much weight.

When analyzing the other galaxies in the group, a notable difference emerges. While the SF regions are also predominantly green, they show bluer hues in the RGB maps, indicating elevated \oiii\ fluxes. These regions appear in the upper-left, low metallicity part of the BPT diagram. This trend is consistent with the mass-metallicity relation, where lower-mass galaxies tend to be poorer in metals. This pattern mirrors what was observed in the M101 group, which also hosts low-mass galaxies and shows a similar extension towards low-metallicity regions in the BPT diagram.
The inclination of NGC 5023 and NGC 5229 prevents a detailed spatial analysis. The irregulars UGC 8320, UGC 8331 and UGC 8308 appear dominated by SF regions in terms of both $W_{\Ha}$ and line ratios. 
UGC 8215, on the other hand, shows weak ELs, despite its young stellar population.

\section{Discussion}\label{sec:discussion}
In this section, we examine scaling relations within the galaxies of the M51 group, highlighting trends, similarities, and differences among them. We also place our results in the broader context of previous findings in the literature.
A key goal of this analysis is to explore how stellar population and EL properties relate to the local stellar mass surface density. Such relations, particularly those linking stellar mass to age, metallicity, and SFR, have been studied in both integrated (e.g., \citealt{Gallazzi2005, Renzini2015}) and spatially resolved data sets (e.g., \citealt{GonzalezDelgado2014MZRletter, GonzalezDelgadoR2016, Sanchez2020_review}).

\begin{figure*}
    \centering
    \includegraphics[width=\linewidth]{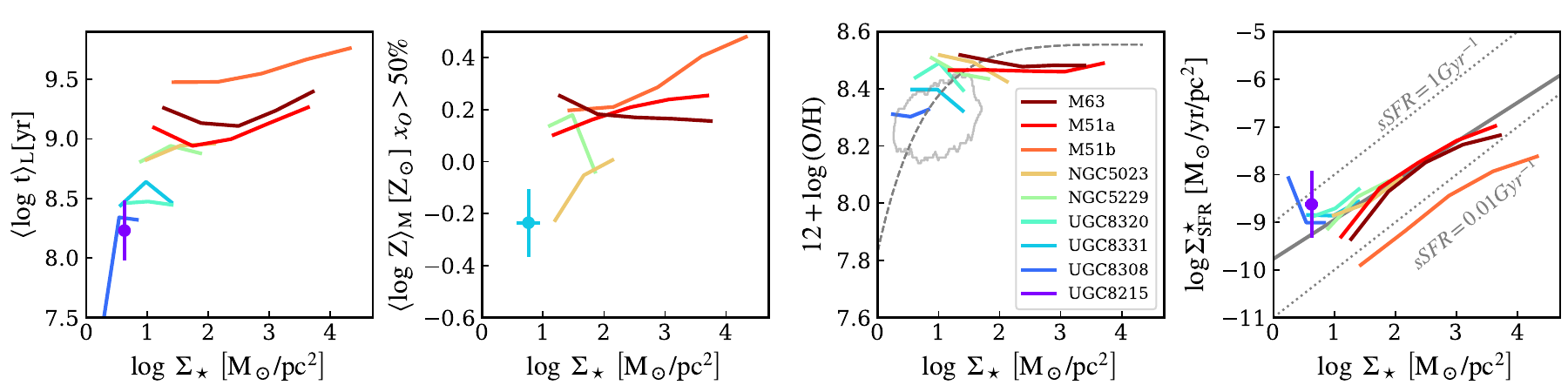}
    \caption{Median scaling relations for all galaxies. In the third column, the gray dashed line is the median curve from \cite{Barrera-Ballesteros2016} for 653 MaNGA galaxies. The gray line shows the 80\% contour of  \cite{Barrera-Ballesteros2016} for MaNGA galaxies with $\log M_\star/M_\odot < 9.2 $.
    The dotted lines in the last panel represent lines of specific SFR at 0.01 and 1 Gyr$^{-1}$. The thick gray line shows the star-forming main sequence relation obtained by \cite{Enia2020_MS} for nearby spirals. Where data are insufficient for a median curve, only uncertainty bars are plotted. 
    }
    \label{fig:mass-relationsM51group}
\end{figure*}

Figure \ref{fig:mass-relationsM51group} shows four relations for our sample: mean stellar age, stellar metallicity, nebular metallicity, and SFR surface density, all plotted against  $\Sigma_{\star}$. As in previous figures, the galaxies are ordered by total stellar mass, but now we have the M51a and M51b analysed separately, which makes M63 the most massive in the group. The lines represent median trends, with bins along the $\Sigma_{\star}$-axis.
As $\Sigma_{\star}$ generally decreases with radius, these plots effectively trace radial trends without 
explicitly referring to a radial coordinate.
Together, these relations offer a spatially resolved view of galaxy structure and evolution, setting the stage for a more detailed discussion in the following subsections.

\begin{figure*}
    \centering
    \includegraphics[width=\linewidth]{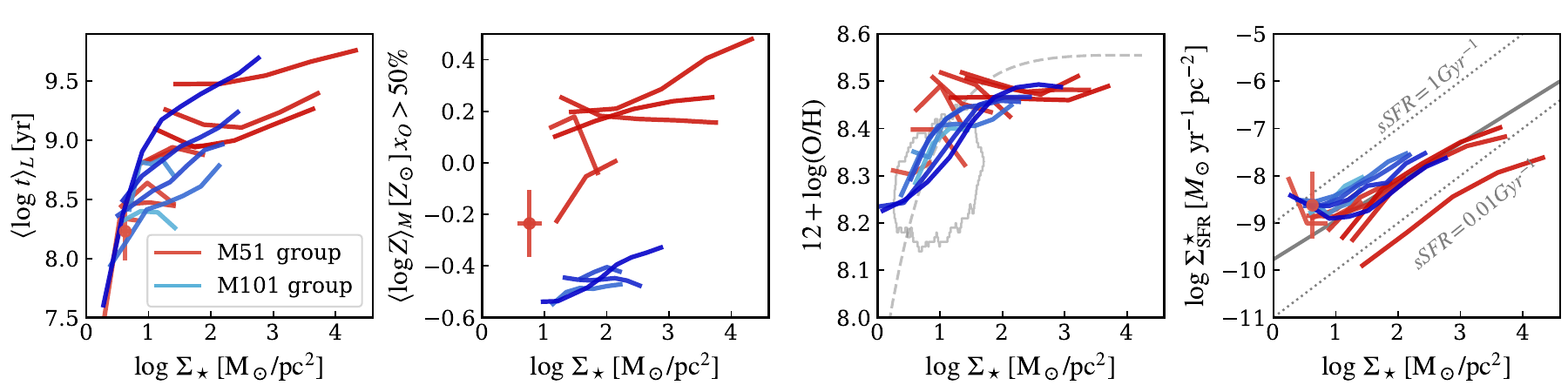}
    \caption{{The panels follow the same structure as in Fig.~\ref{fig:mass-relationsM51group}. Comparison of median scaling relations between the M51 group (red lines) and the M101 group \citepalias[blue lines;][]{ThainaBatista2025}.}
    }
    \label{fig:mass-relationsM51+M101}
\end{figure*}

\subsection{Mass-age relation}\label{sec:age-mass_relationsM51group}

The left panel of Fig.\ \ref{fig:mass-relationsM51group} shows how the mean age indicator, $\langle \log t \rangle_L$, varies as a function of 
$\Sigma_\star$ for galaxies in the M51 group. The first notable effect is the tendency for more massive galaxies to host older stellar populations, a phenomenon commonly referred to as ``downsizing'' \citep{Spitoni2020, PerezGonzalez2008}.

Although M51b is not the most massive galaxy in the sample, it appears at the top of the age-$\Sigma_\star$ relation, consistent with its early-type morphology.
M63 and M51a follow next in the sequence. Next come NGC 5023 and NGC 5229, which have comparable masses and morphologies. Finally, the four low-mass irregular galaxies (UGC 8331, UGC 8320, UGC 8315, and UGC 8308) populate the younger, lower-$\Sigma_\star$ end of the relation. This overall sequence suggests a coherent picture where both stellar mass and morphological type drive the spatially resolved mass-age relation within the group \citep[as discussed in ][]{GarciaBenito2017, GonzalezDelgadoR2015}. At the same time, the causal link may operate in the opposite direction: stellar mass and stellar population age can themselves shape galaxy structure and morphology. Recent results show that age is a strong predictor of galaxy kinematics \citep{Croom2024}, and galaxy kinematics are tightly connected to morphological class through the fast/slow rotator framework \citep{Capellari2016review}. Thus, the mass-age sequence and morphology are likely coupled, with mutual influence rather than a one-way dependence.

Let us now look at the individual galaxies. M51b is the most compact and dense system in the group, reaching values of $\Sigma_\star$ in excess of  $20000 M_\odot/\text{pc}^2$ and a positive mass-age correlation.
M51a also shows a positive age–$\Sigma_\star$ correlation down to $\Sigma_\star \sim 50\ M_\odot,\text{pc}^{-2}$, beyond which the trend reverses, with ages increasing again in the outskirts. We interpret this behavior as a result of the interaction with M51b. This result is consistent with \cite{Wei2021M51}, who found a similar upturn in the radial age profile of M51a around 150 arcsec (which corresponds closely to our $\Sigma_\star \sim 50 / M_\odot \text{pc}^{-2}$ in our map), despite using a different methodology and data.

In M63, we observe a similar trend to that of M51a, a positive age-mass correlation that reverses at $\Sigma_\star \sim 100/M_\odot \text{pc}^{-2}$. However, in the case of M63, the reason for this trend is not so obvious. 
To better interpret this behavior, we must consider the age-mass relation in conjunction with the maps discussed in previous sections, as well as findings from the literature. M63 stellar population and EL maps have a genuine match, showing internal structures of young stars and ionized gas that drops around $3.7$ arcmin (9.2 kpc) radius, becoming dominated by older stellar populations and low \WHa\ values ($\lesssim 10$ \AA).

This observed upturn in the age profile at low $\Sigma_\star$ is likely related to the galaxy's interaction history. M63 exhibits several signatures of past interactions, including a warped \hi\ disk \citep{Battaglia2006M63}, extended tidal streams \citep{Staudaher2015M63}, and evidence of a recent merger with a dwarf companion \citep{Chonis2011M63}. We propose that this minor merger event induced tidal forces that stripped gas from the outer regions of M63. 
Furthermore, the upturn in the age–$\Sigma_\star$ relation could be produced by minor mergers that extend the outskirts of the galaxy. In such a scenario, an increase in stellar ages in the outskirts is expected, since dwarf galaxies are generally old.

As a result, the outer disk is left gas-poor, and the star formation is suppressed. This explains the presence of older stellar populations and the lack of ionized gas in the external parts.
Additionally, this same interaction may account for the asymmetries observed in the \Ha\ emission (as discussed in Sect.\ \ref{sec:results}). Tidal interactions can compress the gas in the disk, enhancing local star formation, particularly along the spiral arms, where we indeed observe concentrated regions of high \WHa.

To further support this interpretation, we computed the radial profile of the sSFR, shown in Fig.\ \ref{fig:sSFR_M63}. The sSFR starts at approximately $10^{-10.7}$~yr$^{-1}$ near the center, rises to about $10^{-10.5}$~yr$^{-1}$ in the intermediate regions, and then drops below $10^{-11}$~yr$^{-1}$ in the outskirts. 
This behavior is consistent with that reported by \cite{Smith2022_M63}, who found a similar trend using the \textsc{CIGALE} code applied to multiwavelength data. These results reinforce the interpretation that M63's outer disk is currently devoid of significant star formation activity, likely as a result of environmental effects driven by a past interaction.

Turning to the less massive galaxies, we observe that they generally do not exhibit clear local mass-age relations. 
We propose that this flatness results from a combination of environmental effects and intrinsic properties related to their low mass. These galaxies are more susceptible to feedback and external perturbations, which can homogenize their stellar populations over time.
Compared to the galaxies in the M101 group, which do show more prominent mass-age gradients, these low-mass systems in the M51 group may be at a more evolved or disturbed stage in their interaction history.

\subsection{Stellar mass-metallicity relation}\label{sec:Z-mass_relationsM51group}

Following the same approach adopted in our study of the M101 group \citepalias{ThainaBatista2025}, to analyze the stellar mass-metallicity relation (MZR), we use the mass-weighted mean (log) stellar metallicity, $\langle \log Z \rangle_M \equiv \sum_j \mu_j \log Z_j$, where $\mu_j$ denotes the mass fraction associated with the component $j$.
For the reasons explained in \citetalias{ThainaBatista2025}, we restrict our analysis of $\langle \log Z \rangle_M$ to regions where populations older than 1 Gyr contribute more than $x_O = 50\%$ of the flux at 5635 \AA. The uncertainties in $\langle \log Z \rangle_M$ (as inferred from the Monte Carlo runs) are of the order of 0.2--0.3 dex for individual zones.

The resulting stellar MZRs are shown in the second column of Fig.\ \ref{fig:mass-relationsM51group}. The scatter around the median curves in these plots is typically $0.2$ dex.
M51b exhibits the highest metallicities, with $\langle \log Z \rangle_M$ reaching values around $3\ Z_\odot$ in the central regions and decreasing to $\sim 1.6\ Z_\odot$ in the outskirts. M51a shows smaller variations in metallicity, with $Z$ decreasing from around $\sim 1.8\ Z_\odot$ in the center to $\sim 1.3\ Z_\odot$ in the outer regions. This same range of values is found for M63, except that in this case the larger metallicities are obtained at the lowest $\Sigma_\star$.

For NGC 5023, NGC 5229, and UGC 8331, we do not observe a well-defined stellar mass-metallicity relation, although their overall lower values of $\langle \log Z \rangle_M$ are consistent with expectations for galaxies with lower stellar masses. In the cases of UGC 8320, UGC 8308, and UGC 8215, no meaningful analysis of the stellar MZR can be performed, as too few spaxels meet the $x_O > 50\%$ criterion required for a reliable estimate of $\langle \log Z \rangle_M$.

\subsection{Nebular mass-metallicity relation}\label{sec:nebOH-mass_relationsM51group}
The third column of Fig.\ \ref{fig:mass-relationsM51group} presents the local nebular MZR. 
Our estimate of the gas-phase metallicity, $12 + \log ({\rm O/H})$, is based on the O3N2 index \citep{Marino2013}, following exactly the same procedure as in \citetalias{ThainaBatista2025}. For individual zones, our estimates of nebular metallicity have Monte Carlo uncertainties of the order of 0.07 dex, which is also the typical statistical scatter around the median curves.

Across the M51 group, the nebular MZR appears generally flat within each galaxy. As in the stellar age-mass relation, we observe a ``downsizing'' pattern: more massive galaxies tend to exhibit higher oxygen abundances, while less massive ones fall systematically below. This trend is well aligned with the median MZR derived by \cite{Barrera-Ballesteros2016} from 653 MaNGA galaxies, shown as a gray dashed line (shifted by $-0.26$ dex to account for IMF differences). The lower-mass galaxies in the group fall within the region defined by their $\log M_\star/M_\odot < 9.2$ sample.

The most massive systems, M51a and M63, show $12 + \log({\rm O/H})$ values around 8.5, consistent with the expected range for Sbc galaxies with $\log M_\star/M_\odot $ of about $ 10.5$-11 as found by \cite{Sanchez2020_review}. \cite{Moustakas2010_M51} got an average oxygen abundance of 8.38 and 8.54 for M63 and M51a respectively, using spectroscopic data.
Our results also agree with those of \cite{Bresolin2004_M51}, who used spectroscopic observations of 10 \hii\ regions in M51a and reported a relatively shallow metallicity gradient compared to other 
spirals (e.g., \citealt{Kennicutt2003_M101}, \citealt{Castellanos2002}). \cite{Croxall2015_M51} analyzed individual \hii\ regions of M51a and confirm 
a small gradient. 
Despite the different methods and calibrations employed in these works, our findings are consistent with their results.

\subsection{Spatially resolved star-forming main sequence}\label{sec:MS_relationsM51group}

The last panel of Fig.\ \ref{fig:mass-relationsM51group} displays the relation between the SFR surface density and stellar mass surface density ($\Sigma^\star_{\rm SFR}$--$\Sigma_\star$), commonly referred to as the spatially resolved SFMS. Most galaxies in the M51 group closely follow the SFMS derived by \cite{Enia2020_MS}, shown as the gray solid line, which was obtained from eight nearby grand-design spiral galaxies. 
We observe that the galaxies align along this sequence with small vertical offsets that correlate with their total stellar mass and morphology, once again reflecting the ``downsizing'' trend identified in previous relations. 
It is also shown that in the low stellar density regions of M63 and M51a, the local SFR lies slightly below the main sequence, indicating that recent star formation has already ceased in the outskirts of these galaxies. See also Fig.\ \ref{fig:sSFR_M63}.

The only significant shift from the SFMS is M51b, which lies well below the sequence, in the region typically associated with quenched or retired galaxies (sSFR $< 0.01$ Gyr$^{-1}$, marked by the diagonal dotted line). This is fully consistent with the previous findings: M51b is the only early-type galaxy in the sample and shows signatures of old stellar populations, weak ELs, and low star formation activity in our maps. 
This result is also supported by the findings of \cite{Nersesian2020_M51}, who reported a lower sSFR for M51b compared to M51a, placing the $\Sigma_\star-\Sigma_{\mathrm{SFR}}$ correlation of the M51b below the quenching threshold of sSFR $=0.01$ Gyr$^{-1}$. Their derived relation for M51a, meanwhile, lies close to the \cite{Enia2020_MS} sequence.

\subsection{Comparison between M51 and M101 groups}

A more detailed comparison between the M51 and M101 groups, including analysis of radial profiles and integrated properties, will be addressed in future work. However, based on the spatially resolved analyses presented here and in \citetalias{ThainaBatista2025}, we can already identify key similarities and differences between the two groups and outline a preliminary comparison.
Both groups are formed by spiral and irregular galaxies and exhibit star formation activity, \Ha\ emission, and asymmetric features.
Notably, both include massive spiral galaxies (M51, M63, and M101) as dominant group members, surrounded by lower-mass satellites. 
However, the group environments and interaction histories differ, which may explain the differences revealed in their spatially resolved stellar population and EL properties.

To perform a direct comparison between the groups, in Fig.\ \ref{fig:mass-relationsM51+M101} we plot the median scaling relations for the M51 group galaxies (red lines) alongside those derived for the M101 group in \citepalias{ThainaBatista2025} (blue lines). This visual juxtaposition reveals clear systematic differences.

Both groups exhibit a clear ``downsizing'' trend in their $\Sigma_*-$age and resolved SFMS relations. 
The groups exhibit different mass ranges, with M51 spanning a wider range in $M_\star$, encompassing both lower and higher masses.
In the $\Sigma_*$-age relation, most M101 galaxies show strong positive gradients (older ages at higher $\Sigma_\star$), whereas in the M51 group, these gradients are generally milder. 
The stellar MZR behaves similarly in both groups: the most massive members display rising $\langle \log Z\rangle_M$ with $\Sigma_\star$, while the low‑mass satellites have essentially flat relations (M63 being a notable exception with a slight upturn at low density).
The absolute stellar metallicity ranges differ: in the M51 group, $\langle \log Z  / Z_\odot \rangle_M$ spans roughly $-0.2$ to $0.5$, whereas in M101 it lies between $\sim -0.55$ and $-0.32$. 
There are also differences in the nebular MZR: all M101 galaxies show a positive O/H-$\Sigma_\star$ gradient, while the M51 members have essentially flat nebular abundances. 
Another key difference is that the M101 group exhibits higher \WHa\ values. 
Conversely, the higher dust attenuation observed in M51a and M63 can be attributed to their more advanced interaction stage.
A final noteworthy point is that the galaxies in the M51 group align closely with the SFMS, showing minimal dispersion. In contrast, the M101 group exhibits offsets relative to the SFMS, shifted toward higher sSFRs. 

We interpret these differences as signatures of distinct evolutionary phases and differences in the group masses. The M101 group appears to be at an earlier interaction stage, preserving its inside-out age and metallicity gradients and sustaining robust star formation. In the M51 group, stronger interactions (notably between M51a and M51b, but also past encounters in M63) have flattened both stellar and nebular gradients, quenched central star formation in M51b, and maybe induced dust production. 

Moreover, the environmental processes have more efficacy as the mass of the group or cluster increases (e.g., \citealt{Alonso2012}).
Our analysis shows that the total stellar mass of the M51 group is around $ 1.0\times10^{11} M_\odot$, while the M101 group is a relatively low mass system with $\sim 2.5 \times 10^{10} M_\odot$. 
This mass difference may help explain why the M51 group shows signs of suppressed star formation (e.g., in M51b, UGC 8215, and the outer regions of M63 and M51a), whereas the M101 group exhibits no substantial evidence of environmental quenching.
These findings are {qualitatively consistent} with the results of \cite{GonzalezDelgado2022} on the effect of the total group stellar mass on the excess of quenched galaxies relative to the field.
They find that for a fixed $M_\star$, the quenched fraction excess in low mass groups is significantly lower than in more massive structures.
Their low-mass group regime is defined as those with $M_{\rm group} < 5 \times 10^{11}$~$M_\odot$, which includes both the M101 and M51 groups.
However, with M51 closer to the upper boundary of this regime, its environment is more susceptible to quenching.
According to their results, the masses of the M101 and M51 groups correspond to a predicted difference in quenched galaxy excess of 0.2 and 0.4, respectively, indicating an agreement with the observed level of environmental impact in our sample. However, with a sample of only two groups, these comparisons should be regarded as indicative rather than conclusive: differences between the two groups may also arise from intrinsic scatter in the scaling relations, the distinct dynamical states of the systems, or assembly bias (e.g., earlier collapse of the M51 group than M101). Notably, the outside-in quenching signatures observed in M51a and M63 favor an environmental contribution, but our sample size prevents us from isolating group mass as the sole driver of these trends.

\section{Summary and conclusions}\label{sec:conclusions}
In this work, we have explored a spatially resolved analysis of the M51 galaxy group using 12-band photometric datacubes from the J-PLUS survey, with the \textsc{AlStar} full spectral fitting code. This study continues the methodology developed for the M101 group \citepalias{ThainaBatista2025}, enabling a consistent comparison between two nearby galaxy groups at very similar distances but different evolutionary/dynamical stages.
Our main findings can be summarized as follows:

\begin{enumerate}
    \item 
    The M51 system shows a dichotomy. M51a is characterized by prominent spiral arms with ongoing star formation. In contrast, its companion M51b is dominated by older, quiescent stellar populations, with an sSFR below 0.01 Gyr$^{-1}$, consistent with a retired/quenched galaxy.

    \item M63 displays significant asymmetries in its stellar age, dust, and $W_{\Ha}$ maps. We find evidence for outside-in quenching, where the outer disk is dominated by older stellar populations and lacks significant star formation. These features are consistent with its past interaction with a dwarf \citep{Chonis2011M63}.

    \item The low-mass galaxies in the group generally show flatter gradients and are dominated by younger stellar populations, consistent with expectations for dwarf irregular and spiral systems.
\end{enumerate}

By comparing the M51 and M101 groups, we conclude that both local (e.g., stellar mass, morphology) and environmental (e.g., group interactions, total mass) factors play decisive roles in shaping galaxy evolution. 

We interpret these results as clear evidence of environmental processing. The more massive nature of the M51 group, combined with its more intense and mature interactions, has significantly altered its members. These processes have flattened chemical gradients, triggered substantial dust production, and initiated quenching processes in the outskirts of M63 and M51a. In contrast, the less massive and dynamically younger M101 group has preserved the primordial inside-out formation signatures of its galaxies.

This comparative analysis demonstrates that the group environment, even at these relatively low densities, plays a crucial role in shaping the evolutionary paths of galaxies. This study further reinforces the capability of multi-band photometric surveys like J-PLUS to drive powerful, IFS-like science over a wide FoV.
In future work, we will expand our sample, adding other groups and clusters, building a better statistical analysis with several dynamic stages. 
Then, expand our analysis to radial profiles and integrated properties to the whole sample to further quantify the environmental effects on galaxy evolution in group environments.

\begin{acknowledgements}
     This work was supported by CAPES under grant 88881.892595/2023-01 and FAPESC (CP 48/2021). 
    RCF acknowledges support from CNPq (grants 302270/2018-3 and 404238/2021-1).
    JTB, RGD, RGB, JRM, GMS and LADG acknowledge financial support from the Severo Ochoa grant CEX2021-001131-S funded by MCIN/AEI/ 10.13039/501100011033 and to grant PID2022-141755NB-I00.
    DRD acknowledges support from CNPq (grant 313040/2022-2).
Based on observations made with the JAST80 telescope and T80Cam camera for the J-PLUS project at the Observatorio Astrof\'{\i}sico de Javalambre (OAJ), in Teruel, owned,
managed, and operated by the Centro de Estudios de F\'{\i}sica del Cosmos de Arag\'on (CEFCA). We acknowledge the OAJ Data Processing and Archiving Unit (UPAD;
\citealt{upad}) for reducing the OAJ data used in this work.
Funding for the J-PLUS Project has been provided by the Governments of Spain and
Arag\'on through the Fondo de Inversiones de Teruel; the Aragonese Government through
the Research Groups E96, E103, E16\_17R, E16\_20R, and E16\_23R; the Spanish Ministry
of Science and Innovation (MCIN/AEI/10.13039/501100011033 y FEDER, Una manera de
hacer Europa) with grants PID2021-124918NB-C41, PID2021-124918NB-C42,
PID2021-124918NA-C43, and PID2021-124918NB-C44; the Spanish Ministry of Science,
Innovation and Universities (MCIU/AEI/FEDER, UE) with grants PGC2018-097585-B-C21
and PGC2018-097585-B-C22; the Spanish Ministry of Economy and Competitiveness
(MINECO) under AYA2015-66211-C2-1-P, AYA2015-66211-C2-2, AYA2012-30789, and
ICTS-2009-14; and European FEDER funding (FCDD10-4E-867, FCDD13-4E-2685). The
Brazilian agencies FINEP, FAPESP, and the National Observatory of Brazil have also
contributed to this Project.
\end{acknowledgements}


\bibliographystyle{aa}  
\bibliography{References2024JT.bib}

\begin{appendix}
\onecolumn
\section{Complementary material}
This appendix presents complementary figures that support the results discussed in the main text.

\begin{figure*}[h]
    \centering
    \includegraphics[width=\textwidth]{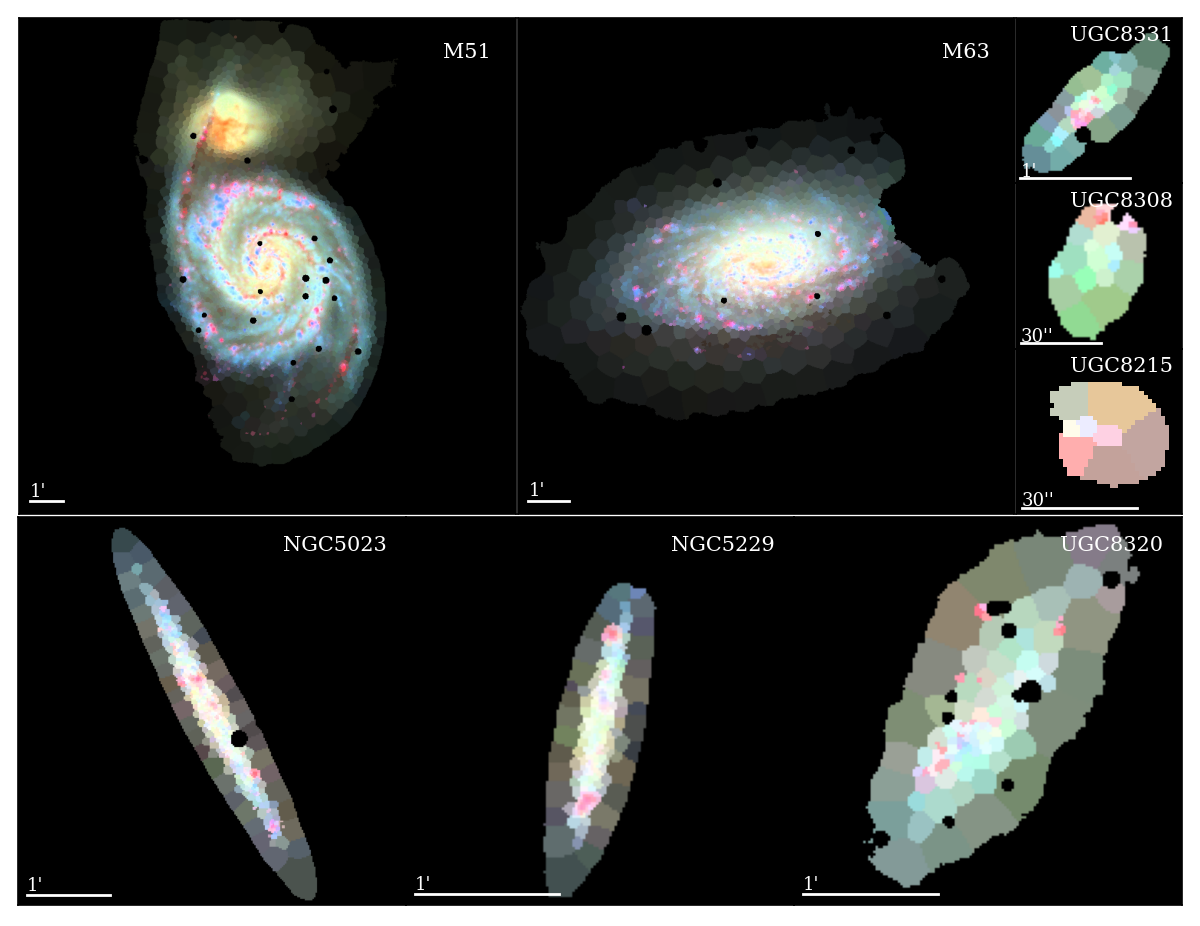}
    \caption{RGB composites using ($J0660$, $g$, sum of five bluer filters) of the pre-processed data to the galaxies of M51 group. The white line represents the arcmin distance.}
    \label{fig:groupM51_finalRGB}
\end{figure*}

~

\begin{figure*}
    \centering
    \sidecaption
\includegraphics[width=0.35\textwidth]{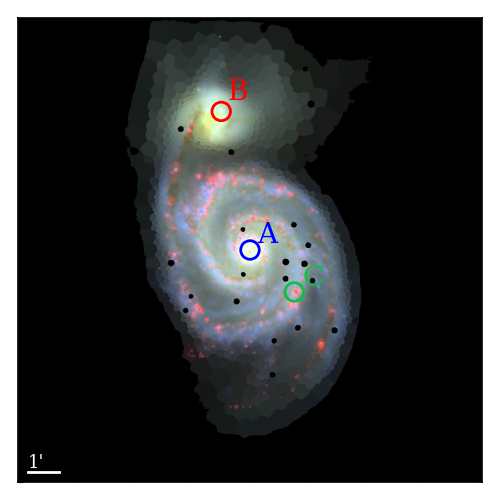}
\includegraphics[width=0.35\textwidth]{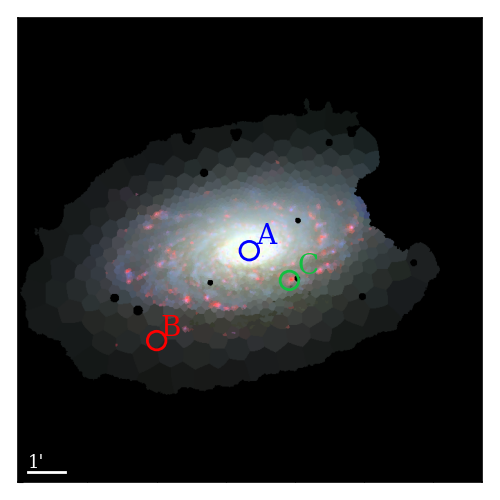}
    \caption{Fit positions (A, B, and C) from Fig. \ref{fig:fit_M51example} in M51 and M63. The RGB composite is the same as in Fig. \ref{fig:fit_M51example}.}
    \label{fig:positions_fitsM51eM63}
\end{figure*}

\twocolumn

\begin{figure*}
\sidecaption
\centering
\vbox{
    \hbox{\includegraphics[width=12cm]{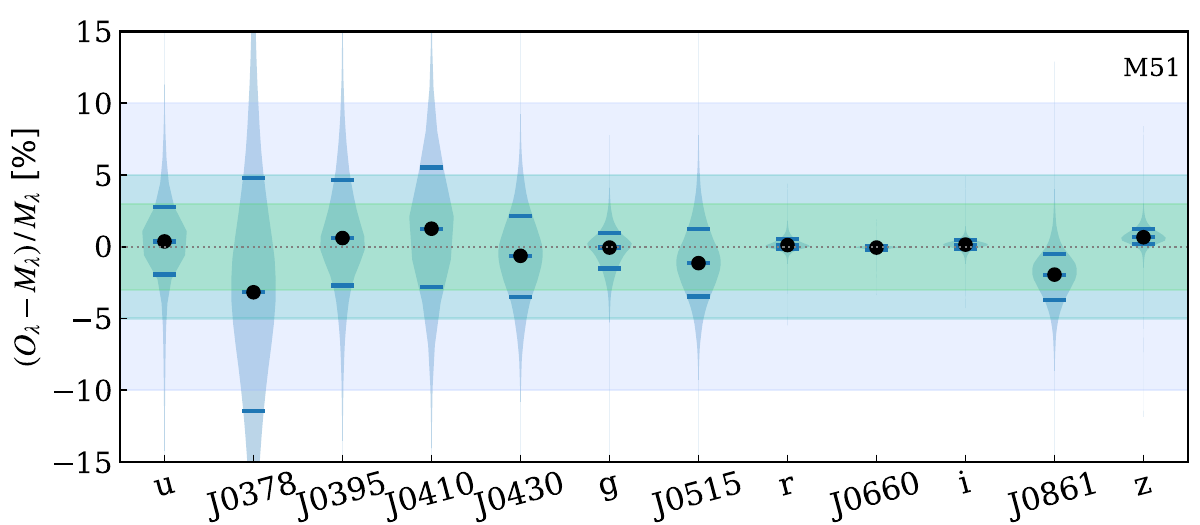}}
    \vspace{0.2cm}
    \hbox{\includegraphics[width=12cm]{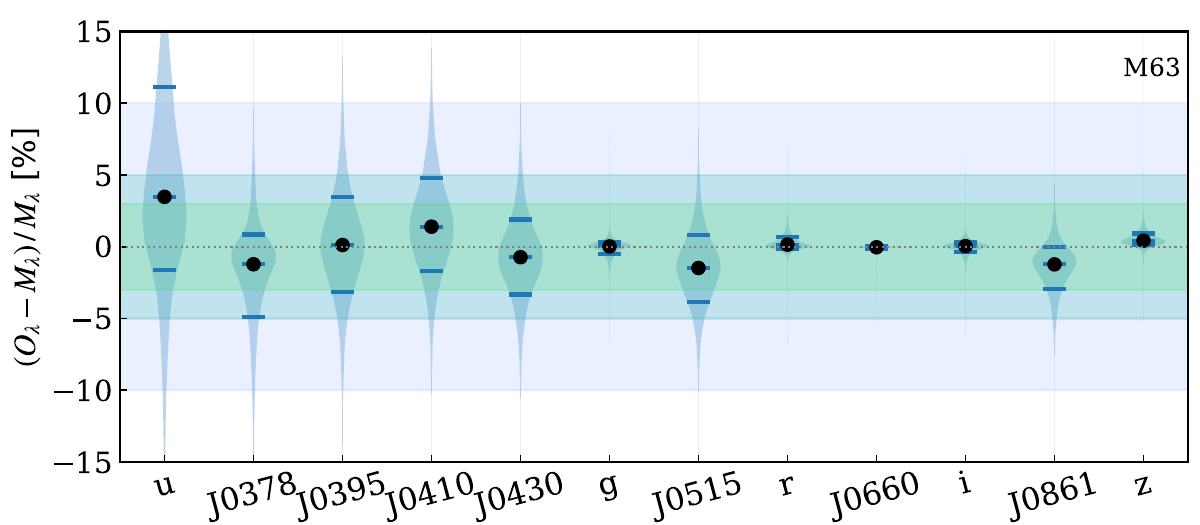}}
}
\caption{Distributions of the $(O_\lambda - M_\lambda) / M_\lambda$ relative residuals of the fits for the 12 J-PLUS bands for zones of M51 and M63. Median residuals are marked by solid black circles, while the horizontal bars mark the 16 and 84 percentiles.}
\label{fig:adev_violin_M51_63}
\end{figure*}

\begin{figure*}
\sidecaption
    \includegraphics[width=12cm]{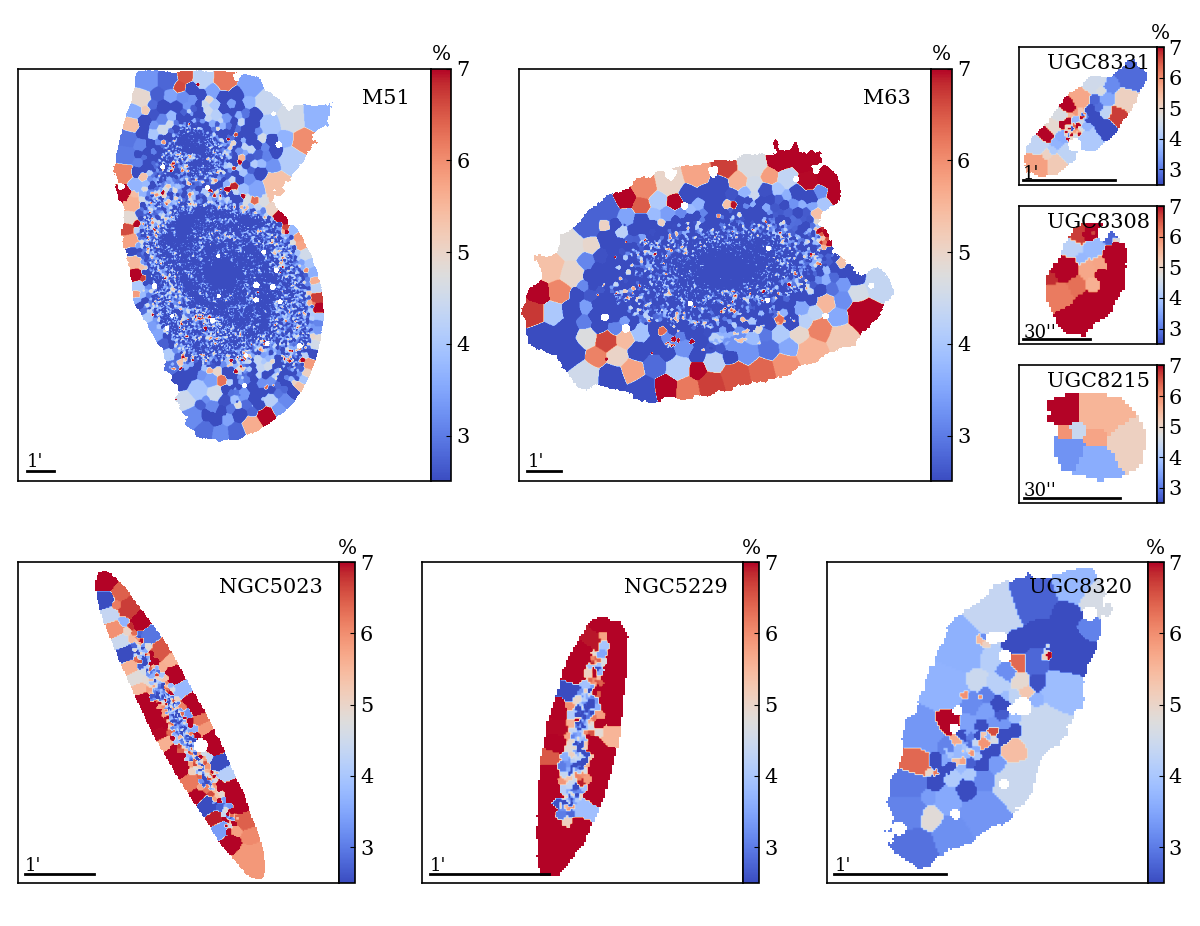}
    \caption{Maps of the mean relative absolute deviation between observed and model fluxes, $\overline{\Delta}$,  to the M51 group. The black line represents the distance.}
    \label{fig:adev_maps_groupM51}
\end{figure*}

\begin{figure}[h!]
    \centering
    \includegraphics[width=0.5\textwidth]{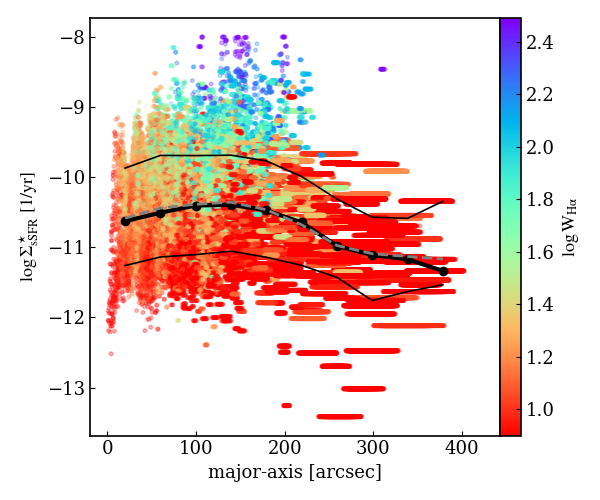}
    \caption{Radial profile of the specific SFR for the M63. The line with circle markers indicates the median curves for bins in the major-axis; the gray dashed line is the mean, and the black thin lines show the corresponding 16th and 84th percentile curves. The points are colored by the $\log W_{\rm H\alpha}$. }
    \label{fig:sSFR_M63}
\end{figure}

\begin{figure*}
    \centering
    \includegraphics[width=\linewidth]{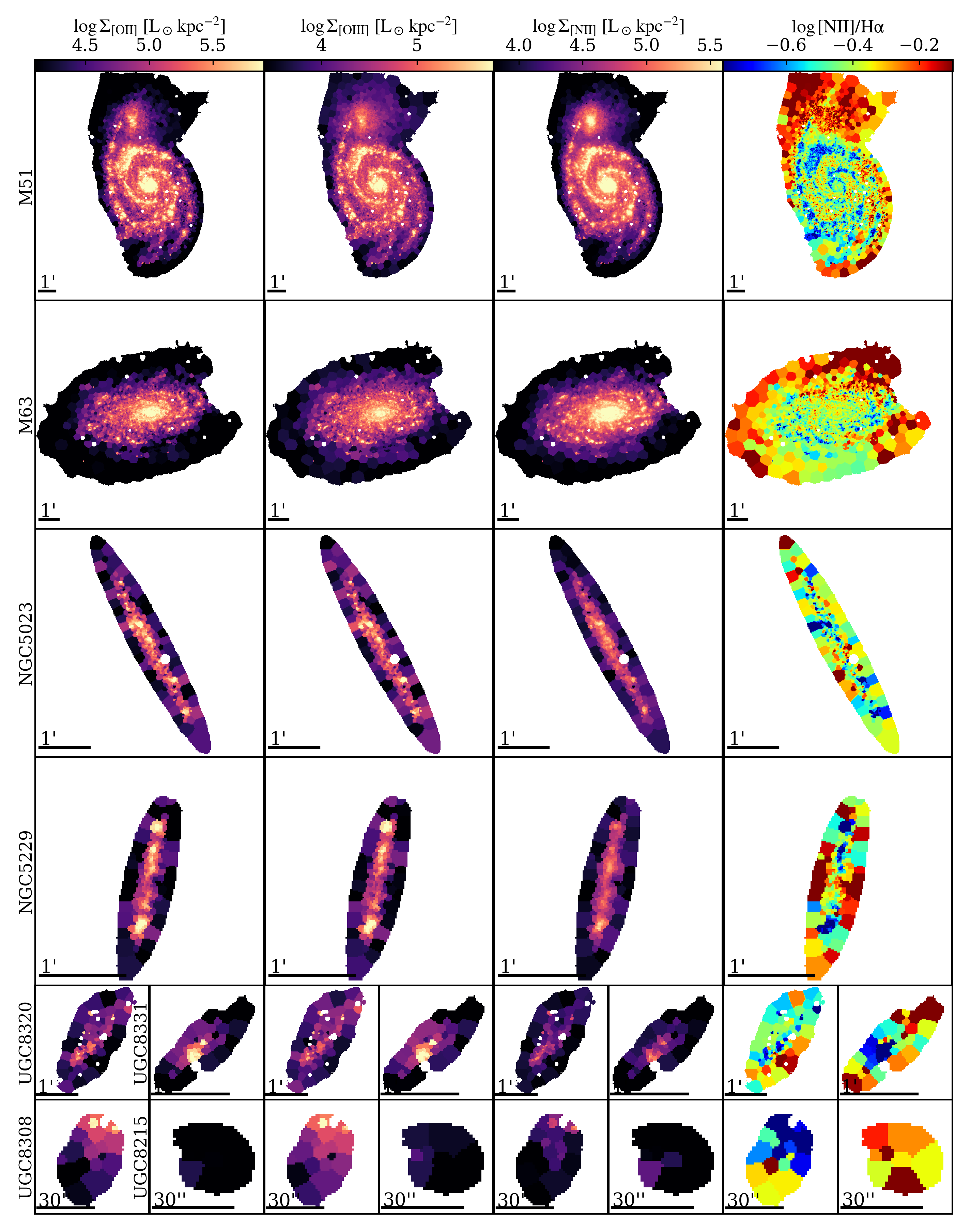}
    \caption{From left to right: Maps of the \oii3727, \oiii5007, and \nii6584 surface brightness, and the \nii/\Ha\ ratio.}
    \label{fig:ELs_extrasM51group}
\end{figure*}

\begin{figure*}
    \centering
    \includegraphics[width=\linewidth]{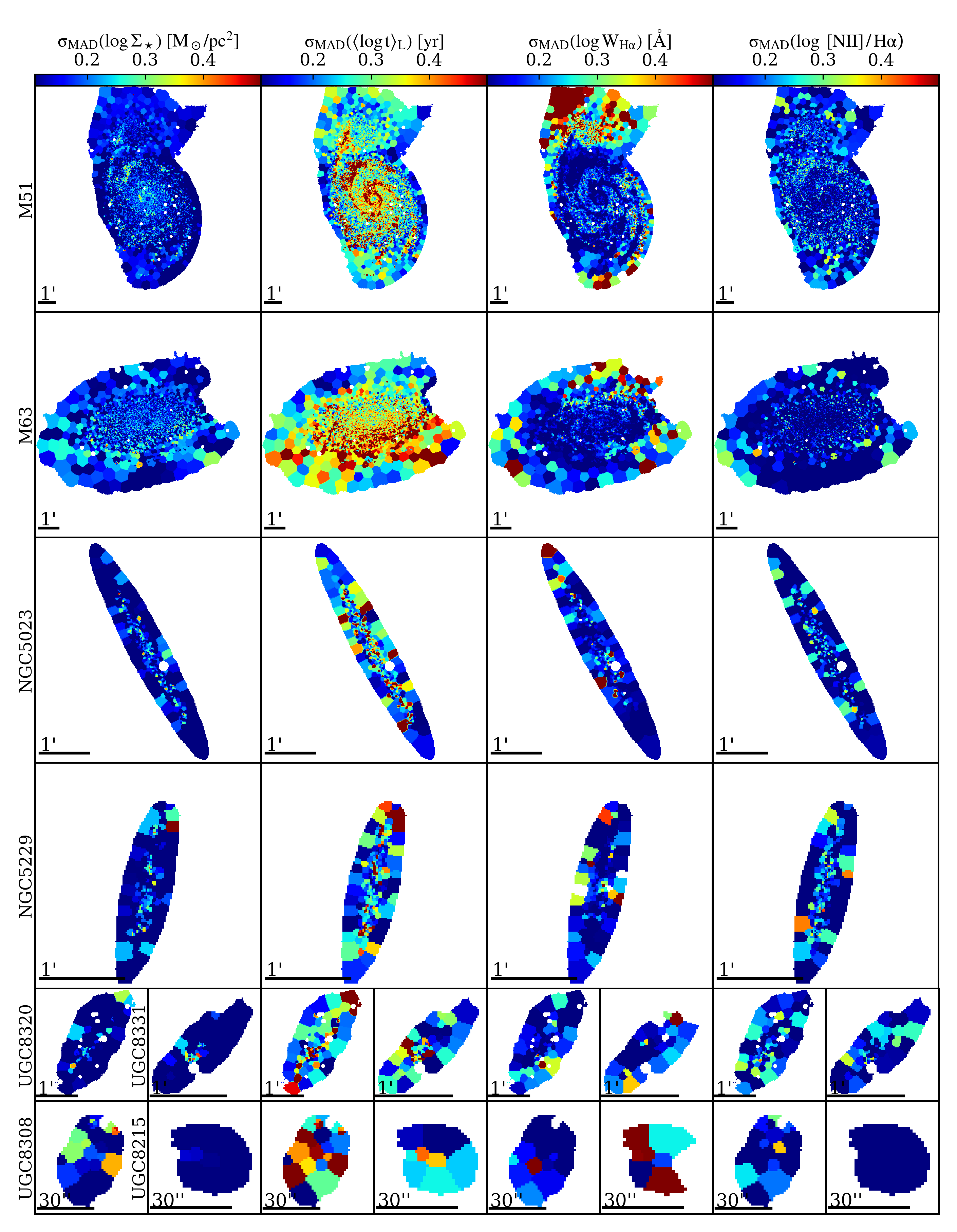}
    \caption{Maps of the $\sigma_{\rm MAD}$ dispersion among the Monte Carlo runs for $\log \Sigma_\star$, $\langle \log t \rangle_L$, $\log \Sigma_{H\alpha}$, $\log W_{\Ha}$ and the \nii/\Ha\ ratio. Where $\sigma_{\rm NMAD}(x) = 1.4826 \times {\rm median}( | x - {\rm median}( x ) | )$.}
    \label{fig:uncertainties_maps}
\end{figure*}

\end{appendix}

\end{document}
